\documentclass[preprint,12pt]{elsarticle}

\usepackage{amsmath,graphicx,float,longtable,url,amssymb,caption,color,dblfloatfix,lineno,hyperref}

\usepackage[ruled]{algorithm2e}
\usepackage{subfigure}
\usepackage{float}
\usepackage[english]{babel}
\usepackage{lipsum}

\modulolinenumbers[5]

\graphicspath{{}}
\def\bx{{\mathbf x}}

\def\by{{\mathbf y}}
\def\br{{\mathbf r}}

\def\bg{{\mathbf g}}
\def\bX{{\mathbf X}}
\def\by{{\mathbf y}}
\def\bY{{\mathbf Y}}

\def\bR{{\mathbb R}}

\def\bzeta{{\mathbf{\zeta}}}

\def\bL{{\mathbf L}}

\def\g{{\mathbf g}}
\def\br{{\mathbf r}}
\def\bq{{\mathbf q}}

\def\bc{{\mathbf c}}

\journal{Digital Signal Processing}

\usepackage{soul}

\begin{document}

\begin{frontmatter}



\title{A Bayesian Particle Filtering Method For Brain Source Localisation}

\author{Xi Chen, Simo S\"{a}rkk\"{a}, Simon Godsill}

\address{Signal Processing Group, Dept. of Engineering, University of Cambridge \\
Department of Biomedical Engineering and Computational Science, Aalto University}

\begin{abstract}
In this paper, we explore the multiple source localisation problem in the cerebral cortex using magnetoencephalography (MEG) data. We model neural currents as point-wise dipolar sources which dynamically evolve over time, then model dipole dynamics using a probabilistic state space model in which dipole locations are strictly constrained to lie within the cortex. Based on the proposed models, we develop a Bayesian particle filtering algorithm for localisation of both known and unknown numbers of dipoles. The algorithm consists of a region of interest (ROI) estimation step for initial dipole number estimation, a Gibbs multiple particle filter (GMPF) step for individual dipole state estimation, and a selection criterion step for selecting the final estimates. The estimated results from the ROI estimation are used to adaptively adjust particle filter's sample size to reduce the overall computational cost. The proposed models and the algorithm are tested in numerical experiments. Results are compared with existing particle filtering methods. The numerical results show that the proposed methods can achieve improved performance metrics in terms of dipole number estimation and dipole localisation.
\end{abstract}


\end{frontmatter}

\section{Introduction}

In recent years, the development of non-invasive brain signal measuring techniques such as MEG and electroencephalography (EEG) have seen rapid progress. These techniques are helpful in diagnosis of mental diseases such as epilepsy, Alzheimer's and Parkinson's disease~\cite{Baillet2001,Kaipio2005}. In non-invasive brain signal processing, we are particularly interested in the signal generated from the cerebral cortex which is the outer layer of the cerebrum~\cite{hamalainen1993,hansen2010meg}. Cortical activity in different cortical regions (such as somatosensory, visual, motor or auditory cortex) can be elicited by suitable stimuli (such as an image or a piece of song). A single active neuron is too weak to be measured directly, so tens of thousands of synchronously active neurons are needed to produce a measurable brain signal. For modelling purposes, many spatially neighbouring active neurons can be summarized and modelled as a dipolar current source, which can be simply named as a ``dipole''. The electromagnetic field generated by such a dipolar source is measurable using MEG/EEG devices. 

Brain source localisation is fundamentally an ill-posed inverse problem~\cite{Kaipio2005,hansen2010meg}. The main barrier is that there may exist many possible solutions for the same set of data, and hence no unique solution can be obtained in the general case. In this paper, we aim to accurately localise the spatio-temporal brain sources using the electromagnetic signals collected outside the surface of the head, employing physiological constraints and soft prior information to regularise the undetermined problem.   

\subsection{Related work}
Brain source localisation is an active research field where a significant amount of work has been done in the past two decades (see, e.g., \cite{hamalainen1993,hansen2010meg, Pascual2002, Long2011,Jun2005, Galka2004,Somersalo2003,auranen2005bayesian,nummenmaa2007hierarchical,Campi2008,Sorrentino2009,Miao2013,Sorrentino2013} and the references therein).

There are two main types of methods: distributed source approaches, and point-wise dipole approaches~\cite{hamalainen1993}. Distributed source methods identify the potential active brain sources that are distributed on a dense grid of fixed locations throughout the whole cerebral cortex (or the whole brain volume if under a looser constraint). Since the number of unknown sources is larger than the number of the M/EEG sensors, mathematical assumptions or constraints are required for an unique solution. Some existing methods include the least squares minimum norm estimation (MNE) \cite{hamalainen1993}, dynamic statistical parametric mapping (dSPM)~\cite{Dale2000}, standardized low-resolution electromagnetic tomography (sLORETA)~\cite{Pascual2002}, and Kalman filter related approaches~\cite{Galka2004,Long2011}.

On the other hand, point-wise dipole approaches treat the brain currents as point dipole sources, and estimate the states (this may include dipole location, moment, and orientation) of the point source dipoles. In this type of modelling, the state of each dipole source is treated as a random unknown target. A number of works have  been published under this type of modelling; these include multiple signal classification (MUSIC) related approaches~\cite{Mosher1999}, Markov chain Monte Carlo related approaches~\cite{Schmidt1999,Jun2005}, and sequential Monte Carlo (or particle filtering) related approaches~\cite{Somersalo2003,Campi2008,Sorrentino2009,Sorrentino2013,Miao2013}. 

Among the various methods proposed, Bayesian particle filtering seems one of the most promising methods for tackling the source localisation problem. In this paper, we develop a point-wise dipolar source localisation approach using Bayesian particle filtering.

\subsection{Bayesian particle filtering methods for dipole localisation problem}

Particle filtering methods have been developed for this application over the last decade. Somersalo et al.~\cite{Somersalo2003} applied a sequential importance resampling (SIR) particle filter for the dipole localisation problem using artificial planar/3D geometry. Results of a two-dipole localisation example was shown using an ideal spherical head model. Campi et al.~\cite{Campi2008} proposed a Rao-Blackwellised particle filter (RBPF) for dipole tracking with single dipole and two dipole examples. It was shown in that work that the RBPF provided better localisation results with lower computational cost than those from a standard particle filter. Sorrentino  et al.~\cite{Sorrentino2009} integrated a random finite set scheme into the particle filter. The method was able to track a time-varying number of dipoles with the maximum dipole number specified in advance. 

Recently, Sorrentino et. al.~\cite{Sorrentino2013} suggested to model the problem using a static dipole setup. The work employed a resample-move particle filter to recursively estimate the dipole moment. Chen et al.~\cite{Chen2013A, Chen2013B} integrated an MNE step into a multiple particle filter method to localise an unknown number of dipoles. The estimation of the dipole number relied on both the MNE step and the previous localisation history. Miao et al.~\cite{Miao2013} also adopted a multiple particle filter method to localise multiple dipoles, using a probability hypothesis density (PHD) filter to perform the estimation of the unknown/time-varying dipole number. The algorithm was implemented and assessed in a real-time field-programmable gate array (FPGA) board. However, it modelled the brain under the ideal spherical head model, which cannot provide a realistic  description of the true human brain.  

\subsection{Our work}
In this paper, we propose a Gibbs multiple particle filtering (GMPF) algorithm for the multiple dipole source localisation problem. The work is developed based on our previous work~\cite{Chen2013A, Chen2013B}. The contribution of this work is described as follows.

Firstly, a continuous head model which forces the state dynamics to strictly remain on the cerebral cortex is developed. To fit with real world applications, we adopt a 1-layer realistic head model, the Nolte model~\cite{Nolte2003}. Although this head model is quite realistic, the off-the-shelf software implementations of it can only be used to evaluate the model at a discrete set of points (the mesh nodes). For distributed source implementations this is all that is needed. However, in our case we need a smooth manifold which defines the cortex surface and hence the discrete set of points is not enough. For this purpose, we adopt a nearest-neighbor (NN) interpolation method to form an approximate continuous cortical manifold. This allows us to formulate the particle filter state directly in terms of the location on the continuous cortex surface.

Secondly, we develop a particle filtering algorithm by integrating a Gibbs sampling iteration step into a multiple particle filtering (MPF)~\cite{Chen2013A} algorithm. Instead of running each component of the MPF only once at each time step, the GMPF iteratively runs the individual components, conditional on the state of the remaining sources, until the state samples converge. This enables the MPF to iterate to obtain a stable state estimate prior to entering the next time step. 

Thirdly, we develop a dipole number dynamic model along with the GMPF method~\cite{Bugallo2007, Mihaylova2012} for localisation with an unknown dynamic number of dipoles. The model generates three potential dipole number predictions based on the estimate from the previous time step. All three predictions are examined and their corresponding state estimates are calculated. A selection criterion is then applied to obtain the optimal prediction results in each time step. Although approximate in a Bayesian sense, this approach improves the accuracy in estimating the number of dipoles, and thereby improves the overall localisation performance of GMPF.

Finally, we apply a computationally adaptive scheme to adjust the number of particles and the state transition range at each step of the algorithm run. In order to generate candidate numbers of sources at each time step, we integrate a standard noise normalized MNE method~\cite{hamalainen1993} and a spatial clustering method~\cite{Gowda1978} to gain some knowledge on the potential dipolar sources. These prior information are used to evaluate the localisation accuracy. We could then adjust the particle size and the state dynamic space in the next algorithm run. 

The remainder of the paper is organized as follows. Section 2 introduces the data modelling procedure. A discrete / continuous head model, a dipole state transition model, and a dipole number dynamic model are described in this section. The localisation algorithm is proposed in Section 3. Both the models and the algorithms are evaluated in Section 4. Section 5 concludes the article.

\section{Data model} \label{DataModel}
We consider a clinical application using an MEG system with $M = 204$ magnetometers -- the proposed method can be applied to other M/EEG settings with slight modifications. Here we use the $204$-sensor MEG application as an example. All the sensors are placed outside the brain surface to obtain non-invasive measurements. We are interested to infer the neural activities within the brain cortical region. The state space is constrained to lie within the cerebral cortex and is denoted as $\Omega$.

For MEG data, a 1-layer realistic head model is introduced to generate the lead-field matrix (the forward matrix), based on a total of \textit{G} fixed vertices on the cortex. An NN (nearest neighbour) interpolation method is used to interpolate the locations between these vertices.    

As described above, the head model comprises $G$ vertices, $\{\bg_1 \cdots \bg_{\nu} \cdots \bg_G\}$; and $F$ triangular faces on the surface of the cortex, created assuming a 1-shell Nolte model for MEG. The width of the head model is 136 mm, and the distance between two adjacent vertices varies between 2.3 mm to 8.4 mm. The lead-field matrix $\bL$ was generated using the statistical parametric mapping (SPM) software~\cite{Friston1994}. Although $\bL$ provides a relatively accurate approximation for the source distribution in the cortical space, it is discretised artificially to a limited number of fixed-location points. We first introduce the traditional discrete real head model using $\bL$. The neural current density is, by contrast, in reality a continuous spatial flow. For this reason, we then propose an interpolated realistic head model for continuous point-wise dipole localisation.  

\begin{figure}[htb]
\centering
\subfigure[]{
\includegraphics[scale = 0.3]{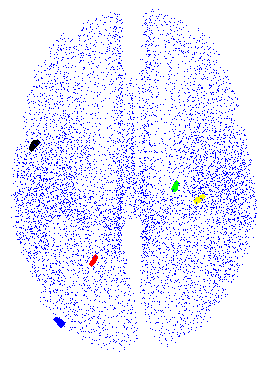}}
\subfigure[]{
\includegraphics[scale = 0.3]{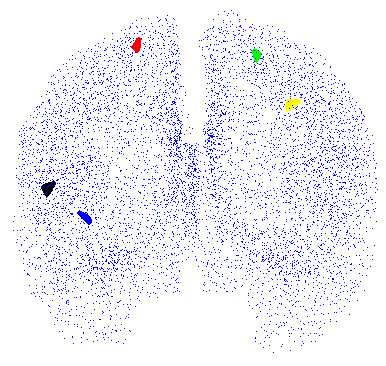}}
\subfigure[]{
\includegraphics[scale = 0.3]{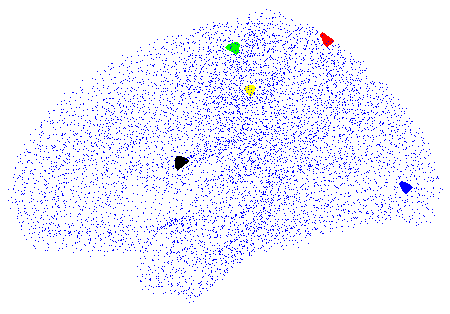}}
\caption{The vertices (the blue dots) and the individual triangular faces (the five small coloured regions) in the real head model.}
\label{BrainCortex}
\end{figure}

Figure \ref{BrainCortex} shows the triangulation of the cortex. The blue dots are the pre-defined vertices on the cerebral cortex, and the 5 coloured small areas are example sub-planes that represent the individual triangular faces on the cortex. In order to better fit real world applications, we strictly enforce that the trajectory of a point-wise dipolar source lies within the modelled cerebral cortex. Each individual dipolar source may only move within a single triangular cortical region defined by the fixed vertices and the triangular faces. Thus we model each dipole as semi-static within a small spatial volume for the whole observation interval.

For a point-wise dipolar source, we define its state as a vector $\bX_k \in \bR^{(6N) \times 1}$, where $k$ is the time instant and $N_{k}$ is the number of dipoles at that time. Since the time resolution of MEG is in the millisecond scale, the unit of the time step in this paper is set as one millisecond. We define the matrix containing the joint state of all dipoles: 
\begin{align}
\bX_k = [\bx_k^1 \cdots \bx_k^n \cdots \bx_k^{N_{k}}]^T.
\end{align}
Here, each $\bx_k^n$ is a state vector for a single dipole, defined as $\bx_k^n = [\br_k^n, \bq_k^n]^T$, 
where $\br_k^n$ is the 3D location and $\bq_k^n$ is the dipole moment (the dipole amplitude with orientation). The dipole orientation is set as normal to the cortical sub-plane surface, so we can usually take $\bq_k^n=q_k^n$ to be just the scalar amplitude of the dipole once the cortical geometry is specified by the head model.

A general measurement model that describes the relationship between the MEG measurement and the dipole states is defined as:
\begin{align}
\bY_k = {\bf H}_{}(\br_k){\bf q}_k + \bzeta_k,  \label{EqGeneral} 
\end{align}
where $\bY_k \in \bR^{M \times 1}$ is the measurement vector at time $k$, $\bY_k = [\by_k^1 \cdots \by_k^m \cdots \by_k^M]^T$. $\bf H(\cdot)$ represents a linear measurement matrix. ${\bf H}_{}(\br_k){\bf q}_k$ is the general measurement model function, ${}\br_k$ and ${\bf q}_k$ denote the vectors of source locations and amplitudes respectively. $\bzeta_{k}$ is the measurement noise vector, which is assumed to contain independent Gaussian random variables with zero mean and variance $\sigma_{\zeta}^2$. 

\subsection{Discrete head model}

Rather than recomputing, expensively, the general model expression above for each new source location, a discretised head model is generated and interpolated in order to approximate the model on the fine scale of the actual source locations.
In order to achieve this, we first compute the lead field matrix corresponding to \textit{\textit{every }}\emph{}possible discretised grid location $\{\bg_1 \cdots \bg_{\nu} \cdots \bg_G\}$. Responses due to arbitrary off-grid source locations are then computed using a special NN interpolation scheme detailed below.

For the discrete head model, the measurement function is simply the discretised lead-field matrix $\bL$~\cite{hamalainen1993}, an $M \times G$ matrix representing  the linear relation between  dipole sources at all possible discrete grid locations and the measurements. The model can be rewritten, assuming once again a linear structure, as:
\begin{align}
\bY_k = \bL \bX_k^{\bL} + \bzeta_k^{\bL},  \label{EqLeadField} 
\end{align}
where $\bX_k^{\bL} \in \bR^{G \times 1}$ is the vector of amplitudes at the  $G$ fixed location vertices in the cortex, $\bX_k^{\bL} = [q_{k,\bg_1} \cdots q_{k,\bg_{\nu}} \cdots q_{k,\bg_G}]^T$. Here, for the discrete model, the set of dipole locations $\br_k^n$ are pre-specified from the fixed spatial grid $[\bg_1 \cdots \bg_{\nu} \cdots \bg_G]^T$.

\subsection{Continuous head model}

\begin{figure}[htb]
\centering
\includegraphics[width = 0.8\linewidth]{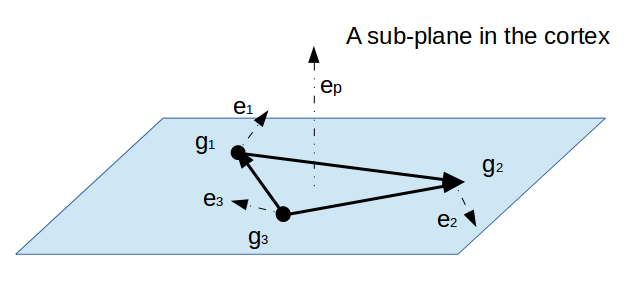}
\caption{The nearest-neighbour interpolation for a three neighbouring vertices example.}
\label{Triangle}
\end{figure}

We apply a simple NN interpolation method~\cite{Coxeter1961} between the $G$ fixed location anchor vertices in order to obtain the approximated continuous head model. As shown in Figure \ref{Triangle}, for the three closest neighbouring anchor vertices to source location $\br_k$ (denoted as $\bg_1$, $\bg_2$ and $\bg_3$ in the example), we obtain $\bY^{\bg_1}$, $\bY^{\bg_2}$ and $\bY^{\bg_3}$. $\bY^{\bg_{\nu}}$ represents the unit noiseless response measured by MEG when we place a unit dipole at the $\nu$th anchor vertex $\bg_{\nu}$, computed using ~\cite{Friston1994}. The area of the triangular region varies depending on the distance between the three anchor vertices, with the average area at around $7$ mm$^2$. The triangular cortical region is sufficiently small that it is reasonable to treat it as a flat sub-plane in which all the interpolated points have the same orientation. In Figure \ref{Triangle}, the orientation that is normal to the sub-plane is denoted as $e_p$. $e_1$, $e_2$ and $e_3$ are the orientations of the three anchor vertices respectively when there is a unit dipole placed at each vertex. We define $\theta$ as the angle between $e_p$ and the orientation of the anchor vertex. We have then $\theta_1$, $\theta_2$ and $\theta_3$ for vertices $\bg_1$, $\bg_2$ and $\bg_3,$ respectively. 

For an $n$th dipole with location $\br^n_k$ at $k$, we  compute its unit response as: 
\begin{align}
\bY^{\br^n_k}= (1-\phi- \varphi) \widetilde{\bY}^{\bg_3} + \phi \widetilde{\bY  }^{\bg_2}  + \varphi \widetilde{\bY}^{\g_1} \label{eq:unit},
\end{align}
where  $\widetilde{\bY}^{\bg_{\nu}} = \cos \theta_{\nu}  \bY^{\bg_{\nu}}$ is the unit response after orientation mapping,  $\phi$ and $\varphi$ are interpolation coefficients  describing the relation between the location of a dipole and the three anchor vertices, computed as follows. 

To obtain the relations between $\phi$, $\varphi$ and the 3D locations $\br^n_k$, $\bg_1$, $\g_2$, and $\g_3$, we define $\alpha = \g_1 - \g_3$, $\beta = \g_2 - \g_3$ and $\gamma = \br_k^n - \g_3$. We have~\cite{Coxeter1961}:
\begin{align}
 \phi &= \frac{(\alpha \cdot \beta)(\gamma \cdot \beta) - (\beta \cdot \beta)(\gamma \cdot \alpha)}{(\alpha \cdot \beta)^2 - (\alpha \cdot \alpha)(\beta \cdot \beta)},\label{eq:interS}\\ 
 \varphi &= \frac{(\alpha \cdot \beta)(\gamma \cdot \alpha) - (\alpha \cdot \alpha)(\gamma \cdot \beta)}{(\alpha \cdot \beta)^2 - (\alpha \cdot \alpha)(\beta \cdot \beta)}.
  \label{eq:interT}
\end{align}

We then obtain the final measurement by summing up all of the scaled unit responses multiplied by its amplitude, from different individual dipoles: $\bY_k = \sum_{n = 1}^{N}  \bY^{\br^n_k}{q}_k^n\simeq{\bf H}_{}(\br_k){\bf q}_k$, as in Equation \ref{EqGeneral}, where the approximation arises as a result of the NN interpolation procedure. The interpolation is executed in the following steps:
\begin{itemize}
 \item Find the triangular sub-planes where the dipoles $\bX_k$ are located in.
 \item For each individual dipole $\bx_k^n$, identify the anchor vertices $\bg_{\nu}$ and compute their corresponding orientations.
 \item Calculate the angles $\theta_{\nu}$ to obtain $\widetilde{\bY}_k^{\bg_{\nu}}$.
 \item Calculate $\phi$ and $\varphi$ using Equation \eqref{eq:interS} and \eqref{eq:interT}, then compute $\bY^{\br^n_k}$.
 \item Sum up all $\bY_k^{\br^n_k}$ to obtain the predicted (noiseless) measurement $\bY_k$.
\end{itemize}

\subsection{Individual dipole dynamic model}

For an individual dipole which exists from time instant $k-1$ to $k$, we define the following individual dipole transition model: 
\begin{align}
\bx_k^n = f(\bx_{k-1}^n, S_{\kappa}^n,\phi^n, \varphi^n),
\label{eq:transition}
\end{align}
where $f(\cdot)$ can be a linear or nonlinear function. $S_{\kappa}$ is the sub-plane, $\kappa$ is the face index and $n$ is the dipole index. Each of the $F$ sub-planes in the 3D cortical space can be treated as a two dimensional plane. Thus we adopt a two dimensional random walk model as our transition function $f(\cdot)$. As we have stated above, the location of a dipole is modelled as semi-static on the cerebral cortex. The state space is constrained accurately on the brain cortical surface $\Omega$, thereby we divide the dipole dynamic into two phases: the transition between different triangular faces and the transition within an individual face. 

For the transition between different faces, we define $\vartheta(\cdot)$ as the neighbouring sub-plane set, which stores all of the neighbouring triangular faces to the face where the dipole $\bx_{k-1}^n$ is located at the previous time step. In practice, this is computed and stored in a lookup table using the grid information provided by the lead-field matrix. 

For the transition within a face, we draw values for the coefficients $\phi^n$ and $\varphi^n$, and randomly select a position within the triangular sub-plane, $\phi^n \sim \mathbb{U}[0,1]$ and $\varphi^n \sim \mathbb{U}[0,1]$ with constraint described in Equation~\eqref{eq:unit}.   

The procedure to perform the dipole transition is as follows:
        \begin{itemize}
 \item For a dipole with state $\bx_{k-1}^n$, find the neighbour sub-plane set $\vartheta(\bx_{k-1}^n)$.
 \item We have $F_\kappa^n + 1$ sub-planes including $F_\kappa^n$ neighbouring sub-planes $S_{\kappa}^n \in \vartheta(\bx_{k-1}^n)$ and the original sub-plane where $\bx_{k-1}^n$ located. We randomly select one sub-plane with equal probability $p_{\kappa}^n = \frac{1}{F_{\kappa}^n + 1}$.
 \item Randomly choose a position using $\phi^n$ and $\varphi^n$ in the selected sub-plane.
 \item Calculate $\bx_k^n = f(\bx_{k-1}^n, S_{\kappa}^n,\phi^n, \varphi^n)$ to obtain the new dipole state prediction. 
 \end{itemize}

$F_\kappa^n$ is the total number of the identified neighbour faces. In the initialisation step, dipole states are randomly selected from the $G$ fixed location grid points. Since a grid point connects several sub-planes, it is difficult to define which sub-plane it belongs to. In practice, we compute the distance between the grid point and all its neighbour sub-planes. We select the one with the shortest distance to the grid point as its sub-plane. 

\subsection{Dipole number dynamic model}
In our model, we assume that the number of the active dipolar sources does not change dramatically between adjacent time steps. For scenarios with more than one dipole currently active, we can model the dipole number transition process by allowing an individual dipole to appear or disappear at a single time instant. 

Given the dipole number $N_{k-1}$ from the previous time instant, the current dipole number $N_k$ can be obtained from the following dipole number dynamic model: 
\begin{align}
 p(N_k \mid N_{k-1}) = \left\{
  \begin{array}{l l}
    p_{k,(+)} & \quad \text{for $N_k^{} = N_{k-1} + 1$}\\
    p_{k,(0)}  & \quad \text{for $N_k^{} = N_{k-1}$}\\
    p_{k,(-)} & \quad \text{for $N_k^{} = N_{k-1}-1 $}
  \end{array} \right.
  \label{dipolenumber}
\end{align}
 with $\sum_j p_{k,j} = 1$, where $j = \{(-),(0),(+)\}$. The dynamic probability $p_{k,j}$ is predefined with a consideration of dipole birth-death movement~\cite{Ng2007}. In general, we set $p_{k,(+)} = p_{b}$, $p_{k,(-)} = \frac{p_{d}}{N_{k-1}}$, $p_{k,(0)} = 1 - \frac{p_{b} - p_{d}}{N_{k-1}}$. $p_b$ and $p_d$ are the birth-death probability discussed as follows. 

\subsubsection{Dipole birth-death move}
We adopt a simple birth-death move for cases when a new dipole appears or an existing dipole disappears from time $k-1$ to $k$. We set $p_b = p_d = \frac{p_i}{2}$, where $p_i \sim \mathbb{U}(0,1)$. For initialisation step or special cases when $N_{k-1} = 0$, we set $p_d = 0$ and $p_b = p_i$.  

For the birth process, a randomly selected initial state within the cortical surface $\Omega$ is assigned for the new birth dipole. The initialisation is assisted by the results from the probabilistic ROI estimation (detailed in Section \ref{MainAlg}). The probabilistic ROI estimation provides us knowledge about the signal strength of the $G$ grid points in $L$. A probabilistic sampling method is used for sample selection regarding the signal strength (explained in Section 3.2). The selected sample point set is defined as $\Psi_{k}$. We randomly select one of the grid points in $\Psi_{k}$ and assign its location to the new birth dipole. $\Psi_{k}$ is also used to generate the ROIs by applying a spatial clustering algorithm~\cite{Gowda1978} to the selected point sets. 

For the dipole death process, the algorithm randomly selects and deletes one of the existing dipoles. The random selection does not introduce extra bias in the death process due to the large sample number and resampling step in the particle filter. Instead of computing a single result in each time step, the particle filter will draw a number of sample points and compute their weights; samples with lower weights are filtered out to improve the estimation accuracy. Therefore the deleted existing dipoles vary from sample to sample, only samples with a higher posterior density will be used for the final estimates. The particle filtering algorithm will be detailed in Section \ref{MainAlg}.    

\section{Bayesian sequential Monte Carlo algorithm} \label{MainAlg}
In this section, we develop a Bayesian particle filter algorithm to deal with the localisation with both known and unknown dipole number. The dipole localisation problem is cast as a general multi-target tracking problem in our Bayesian framework.

\subsection{Algorithm execution}

\begin{figure}[t]
\centering
\includegraphics[width = 1\linewidth]{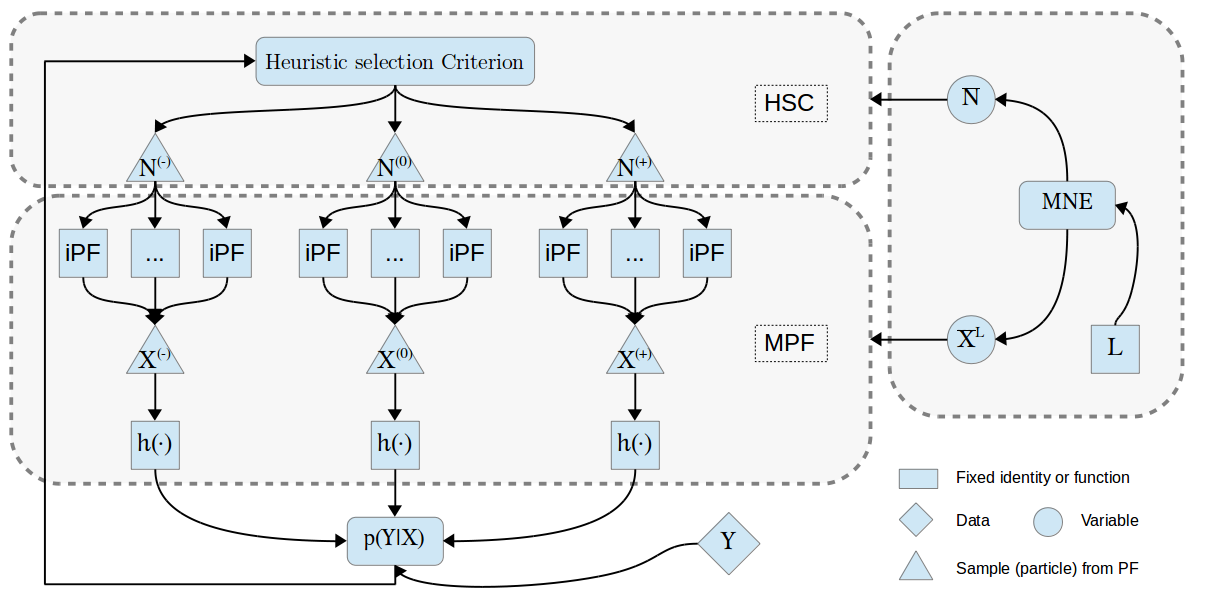}
\caption{Algorithm execution illustration}
\label{Relationship}
\end{figure}

We first describe the algorithm  structure, followed by a detailed description of each component in the proposed algorithm. The algorithm contains three different interactive components: the probabilistic ROI estimation step estimates the number of ROIs $\overline{N}_k$ at time $k$; $\overline{N}_k$ provides an initial guess of the dipole number for the main algorithm; the GMPF step performs the particle filtering; the selection criterion step selects the optimal number $\hat{N}_k$ and its corresponding state estimates $\hat{\bX}_k$ at time $k$. Figure~\ref{Relationship} illustrates the relationship between the components and their corresponding variables. The target number propagation model and the selection criterion scheme are relatively naive and simple when compared to some other existing estimation methods \cite{Ng2007}. In this paper we focus on the tracking algorithm performance. 

The algorithm executes in the following four steps: 
\begin{enumerate}
 \item A probabilistic ROI estimation step uses the noise normalised MNE method and a probabilistic sampling method to obtain the point set $\Psi_{k}$. Therefore we can obtain an estimate for the number of the ROIs $\overline{N}_k$. $\overline{N}_k$ is used for the dipole number initialisation at $N_0$ at $k=0$.  
 
 \item Given the estimated dipole number $\hat{N}_{k-1}$ from $k-1$, three potential dipole numbers $N_k^{j}$  (with their corresponding dynamic probability) can be obtained from Equation \eqref{dipolenumber}.
 
 \item For each $N_k^{j}$, the GMPF algorithm is applied to estimate the dipole state. The algorithm assigns an individual particle filter (iPF) for each of the identified targets. In a single GMPF run, each iPF is executed to generate the estimate $\bx^n_k$ for its corresponding dipole. $\bx^n_k$ is immediately updated in the dipole state $\bX_k^{j}$ to assist other iPFs. This updating procedure is performed in a Gibbs sampling manner (see Section \ref{Gibbs} for details). We  finally obtain the estimated dipole state $\widetilde{\bX}_k^{j}$ for each of the three potential $j$ cases. 
  
 \item The selection criterion scheme is then used to compute the posterior probability for each of the three cases. One of the three cases is selected as the final estimate at $k$. An estimation for both the dipole number $\hat{N}_k$ and the dipole state $\hat{\bX}_k$ can then be obtained. We set $N_{k} = \hat{N}_k$ and $\bX_k = \hat{\bX}_k$. 
\end{enumerate}
In addition, the discrete point set $\Psi_{k}$ from the MNE is used to adaptively control the number of particles and the state transition range in the particle filtering. In the remainder of this section, we give a more detailed description for each parts.

\subsection{Probabilistic ROI estimation } \label{ROISec}
A noise-normalised MNE approach is employed to obtain the signal strength of the fixed grid points. The discrete head model from Equation \eqref{EqLeadField} is $\bY_k = \bL \bX_k^{\bL} + \bzeta_k^{\bL}$. A typical noise-normalised MNE solution~\cite{hamalainen1993} can be obtained as follows:
\begin{align}
 \overline{\bX}_k^{\bL} = \bL^T(\bL \bL^T + \lambda I)^{-1}\bY_k, 
 \label{eq:Prior}
\end{align}
where $\overline{\bX}_k^{\bL}$ is an amplitude estimation vector for the $G$ grid points, $\overline{\bX}_k^{\bL} = [\overline{q}_{k,\bg_1} \cdots \overline{q}_{k,\bg_{\nu}} \cdots \overline{q}_{k,\bg_G}]^T$. $\lambda$ is a noise normalised regularisation parameter. 

We do a simple probabilistic sampling with respect to the estimated amplitude of the $G$ points, as follows:
\begin{align}
 \bg_{k,\nu} \sim \frac{\overline{q}_{k,\nu}}{\sum_{\nu = 1}^{G} \overline{q}_{k,\nu}}.
\end{align}

We now have the point set $\Psi_{k} = \{\bg_{k,1}, \cdots, \bg_{k,\nu}, \cdots, \bg_{k,G} \}$ . We employ a hierarchical spatial clustering method~\cite{Gowda1978} to cluster the selected points in $\Psi_{k}$ with respect to their corresponding geographical positions. We define these clusters as the ROIs. We can obtain from this the number and extent of the ROIs $\Psi_{k}(n)$, $n = 1, 2, \cdots, \overline{N}_k$.

$\Psi_{k}(n) = \{\bg_{k,1} \cdots \bg_{k,\nu^n} \cdots \bg_{k,G^n}\}$ is the subset of the grid points for active region $n$, where $G^n$ is the total number of points in that subset and $\nu^n$ is the grid point index. We compute the location of the $n_k$th ROI by taking the mean of the points in $\Psi_{k}(n)$ so that the centre of the ROI is defined as:
\begin{align}
 \bc^{n}_{k} = \frac{1}{G^n} \sum_{\nu^n = 1}^{G^n} \bg_{k,\nu^n}.
\end{align}
The centre location of each ROI will be used in Section \ref{adaptiveFilter}. 

\subsection{Multiple particle filtering}
For each value of $N_{k}^{j}$, the problem reduces to track a fixed number of dipoles. We target the joint filtering distribution for all sources, $p(\bX_{k}^{j} \mid \bY_{1:k}, N_{1:k-1}, N_k^{j})$, where $N_{1:k-1}$ is the previous estimated dipole number up to  time $k-1$, and $N_k^{j}$ is the current value of dipole number for the $j$th case. The state vector of each GMPF is denoted by $\bX_{k}^{j}$. We assign a GMPF for each $N_{k}^{j}$; the three GMPFs operate in parallel and are independent of each other. 

For each $j$, there are $N_{k}^{j}$ dipoles and therefore $N_{k}^{j}$ iPFs are assigned, one for each active source, as shown in Figure~\ref{Relationship}. The state vector can be expanded as $\bX_{k}^{j} = \{\bx_{k}^{1}, \bx_{k}^{2}, \cdots \bx_{k}^{n_{k}^{j}}, \cdots \bx_{k}^{N_{k}^{j}}\}$, where $\bx_{k}^{n_{k}^{j}}$ denotes the state of the $n_{k}^{j}$th individual dipole out of $N_k^{j}$. To simplify the notation, rewrite the state $\bx_{k}^{n_{k}^{j}}$ as $\bx_{n,k}^{j}$.

Now, defining $\psi_{k}^{j} = \{N_{1:k-1}, N_k^{j}\}$ as the trajectory of the previous and the current active source numbers, the joint target distribution $p(\bX_{k}^{j} \mid \bY_{1:k},\psi_{k}^{j})$ can be expressed as:
\begin{align}
 p(\bX_{k}^{j} \mid \bY_{1:k}, \psi_{k}^{j}) = \frac{p(\bY_{1:k} \mid \bX_{k}^{j}, \psi_{k}^{j}) p(\bX_{k}^{j} \mid \psi_{k}^{j})}{p(\bY_{1:k} \mid  \psi_{k}^{j})}.
\end{align}

As $\psi_{k}^{j}$ is known in the current time step and the dipoles are assumed to be independent of each other \textit{a priori}, we have: 
\begin{align}
 p(\bX_{k}^{j} \mid \bY_{1:k}, \psi_{k}^{j}) \propto p(\bY_{1:k} \mid \bx_{1,k}^{j}, \bx_{2,k}^{j}, \cdots, \bx_{N,k}^{j}, \psi_{k}^{j}) \prod_{n=1}^{N} p(\bx_{n,k}^{j} \mid \psi_{k}^{j}),
\end{align}
where $p(\bx_{n,k}^{j} \mid \psi_{k}^{j})$ is the predictive distribution for each independent source; the states are uniformly drawn from the whole state space $\Omega$. 

In order to draw samples from the joint target posterior distribution, we consider a Gibbs sampling structure~\cite{Gilks1995} to iteratively update the desired dipole states at each time step, in which a conditional particle filter for each active source approximates the required full conditional draw required for a Gibbs sampler. The basic idea is very simple: at time $   k-1 $\ we assume that particles are available from from the joint target $p(\bX_{k-1}^{} \mid \bY_{1:k-1}, \psi_{k-1}^{})$. In order to run this forward one time step, we initialise the state $\bX_{k }$ to some arbitrary initial values. We then run conditional particle filters for each target in turn. Each particle filter is desiged to target the conditional filtering distribution  $p(\bx_{n,k}^{} \mid \bx_{-n,k},\bY_{1:k}, \psi_{k}^{})$, where $\bx_{-n,k}$\ denotes the state $\bX_{k }$ with source number $n$ removed. A single sample is randomly selected from the particle approximation to   $p(\bx_{n,k}^{} \mid \bx_{-n,k},\bY_{1:k}, \psi_{k}^{})$      and this is substituted into the state vector  $\bX_{k }$ at the $n$th source position. One iteration of the Gibbs sampler comprises a complete sweep through all of the particle conditional distributions, and convergence will occur after some large number of iterations (in practice though we implemented just a few iterations). In this way we aim to overcome the approximation induced by the standard MPF, which accounts only approximately for the statistical coupling between sources, at the expense of an iterative procedure at each time step. An interesting modification of our approach would be to incorporate pseudo-marginal sampling ideas into the Gibbs sampler, which would in addition correct  the approximation of the particle filter to the conditional filtering distribution. This addition is left as a future topic for exploration.

We now specify in detail the steps required for the GMPF.

\subsubsection{Gibbs sampling updating step for conditional posterior distribution} \label{Gibbs}
In each time step, the iPFs are executed from the first iPF to the $N_{k}^{j}$th iPF sequentially. The updating procedure acts in a similar way to that in a standard Gibbs sampler: once a new state $\bx_{n,k}^{j}$ is generated, it is immediately used to assist other iPFs by updating the corresponding conditional posterior distribution. In order to obtain a good set of samples from the joint filtering distribution this Gibbs sampler should run for a number of iterations at each time step. We denote the iteration number by the iteration indices $l' = \{1, 2 \cdots l \}$. 

The conditional posterior distribution of the individual dipole state which is approximated by each iPF can be written as $p(\bx_{n,k}^{j(l')} \mid \bY_{1:k}, \psi_{k}^{j}, \bx_{-n,k}^{j(l')})$. The iPF state $\bx_{n,k}^{j(1)}$ is initialised to $\bx_{n,k-1}^{(l)}$ from the previous time step after the selection criterion step. $\bx_{-n,k}^{j(l')}$ describes a state vector excluding the state $\bx_{n,k}^{j}$ at the $l'$th (when $l' \geq 2$) Gibbs iteration:
\begin{align}
 \bx_{-n,k}^{j(l')} = \{ \bx_{1,k}^{j(l')}, \cdots \bx_{n-1,k}^{j(l')}, \bx_{n+1,k}^{j(l'-1)} \cdots \bx_{N,k}^{j(l'-1)} \}.
\end{align}
$\bx_{-n,k}^{j(l')}$ is used in the estimation of the $n$th iPF, its first $n-1$ elements are the estimates updated in the current Gibbs iteration while the other elements are the estimates from the previous Gibbs iteration. Once a new state $\bx_{n,k}^{j(l')}$ is generated, we update the corresponding $\bx_{-n,k}^{j(l')}$ and use it in the $(n+1)$th iPF. The basic scheme of the Gibbs iteration is described as follows:

\begin{itemize}
 \item Generate $\bx_{1,k}^{j(l')}$ from $p(\bx_{1,k}^{j} \mid \bY_{1:k}, \psi_{k}^{j}, \bx_{-1,k}^{j(l')})$. \\
 \item Generate $\bx_{2,k}^{j(l')}$ from $p(\bx_{2,k}^{j} \mid \bY_{1:k}, \psi_{k}^{j}, \bx_{-2,k}^{j(l')})$. \\
 \vdots
 \item Generate $\bx_{N,k}^{j(l')}$ from $p(\bx_{N,k}^{j} \mid \bY_{1:k}, \psi_{k}^{j}, \bx_{-N,k}^{j(l')})$. 
\end{itemize}

Therefore, the final iteration gives $\bX_{k}^{j(l)} = \{\bx_{1,k}^{j(l)}, \bx_{2,k}^{j(l)}, \cdots \bx_{n,k}^{j(l)}, \cdots \bx_{N,k}^{j(l)} \}$, where $\bx_{n,k}^{j(l)}$ denotes the sampled value of a dipole state at time step $k$ for the $j$th case, obtained after $l$ iterations of Gibbs update. After the selection criterion step for each algorithm run, one of the three $j$ cases is selected, we will have $\hat{\bX}_k$ where the notation $j$ is eliminated. This is important because we use some variables at $k-1$ in our Bayesian inference. For example, the term $\psi_{k-1} = \{N_{1:k-2}, N_{k-1} \}$ and $\bx_{-n,k-1} = \{ \bx_{1,k-1}, \bx_{2,k-1}, \bx_{N,k-1} \}$. 

The Gibbs sampling iteration enables the algorithm to get a more accurate estimate in each particle filtering step, particularly when the sources are spatially close and hence not independent in their joint posterior. This raises several issues such as the Gibbs iteration number, convergence analysis, and the computational load considerations. In theory, many iterations would be required to guarantee convergence, but here we only operate a few iterations because of the high computational burden. The intuition here is that usually the dipoles are well separated and quite independent; hence the Gibbs sampling should be approximately converged within a few iterations. 

\subsubsection{Individual particle filtering}
Since $\psi_k^j$ and $\bx_{-n,k}^{j(l')}$ are available terms to each iPF, we only need to  sample the unknown state $\bx_{n,k}^{j(l')}$ in each iPF. The conditional posterior distribution can be rewritten as $p(\bx_{n,k}^{j} \mid \bY_{1:k}, \psi_{k}^{j}, \bx_{-n,k}^{j(l')})$. We define $I_k$ as the number of particles for each iPF at time $k$. A particular sample in an iPF is denoted as $\bx_{n,k}^{ij}$, which is updated in each of the $l'$th Gibbs iteration. 

The conditional posterior distribution required for Gibbs sampling can be expanded in two steps: an updating step and a prediction step.  The (conditional) updating step can be expressed as: 
\begin{align}
 p(\bx_{n,k}^{j} \mid \bY_{1:k}, \psi_{k}^{j},\bx_{-n,k}^{j(l')}) = \frac{p(\bY_{k}^{} \mid \bx_{n,k}^{j}, \psi_{k}^{j},\bx_{-n,k}^{j(l')}) p(\bx_{n,k}^{j} \mid \bY_{1:k-1},\psi_{k}^j,\bx_{-n,k}^{j(l')})} {p(\bY_{k}| \bY_{1:k-1},\psi^j_{k},\bx_{-n,k}^{j(l')})},
\end{align}
where 
\begin{align*}
 &  p(\bx_{n,k}^{j} \mid \bY_{1:k-1},\psi_{k}^j,\bx_{-n,k}^{j(l')}) \\
 & = \int p(\bx_{n,k}^{j} \mid \bx_{n,k-1},\psi^j_{k},\bx_{-n,k}^{j(l')}) p(\bx_{n,k-1} \mid \bY_{1:k-1},\psi^j_{k},\bx_{-n,k}^{j(l')}) d \bx_{n,k-1}
\end{align*}
is the corresponding prediction step. The conditional posterior for $\bx_{n,k-1}$ at $k-1$ is $p(\bx_{n,k-1} \mid \bY_{1:k-1},\psi^j_{k},\bx_{-n,k}^{j(l')})$, containing parameters $\psi^j_{k}$ and $\bx_{-n,k}^{j(l')}$ which are known prior to the iPF run. The conditional posterior distribution $p(\bx_{n,k}^{j} \mid \bY_{1:k}, \psi_{k+1}^{j},\bx_{-n,k+1}^{j(l')}) \approx p(\bx_{n,k}^{j} \mid \bY_{1:k}, \psi_{k}^{j},\bx_{-n,k}^{j(l')})$, which forms a complete recursive form in the Bayesian model. This is because the dipole sources are a priori independent of each other, conditioning on $\psi_{k+1}^j$ and $\bx_{-n,k+1}^{j(l')}$ does not affect the proposed conditional posterior distribution. Here we simply take the weighted samples from the $n$th iPF, building in the independence approximation
\begin{align}
p(\bX_k^j \mid \bY_{1:k},\psi_{k+1}^{j}) \approx \prod_{n=1}^{N_k^j} p(\bx_{n,k}^{j} \mid \bY_{1:k}, \psi_{k+1}^{j},\bx_{-n,k+1}^{j(l')}).
\end{align}

In these equations we have also built in the assumptions that the sources are a priori independent at each time point, and that the observations at time \textit{k} are conditionally independent of the states prior to time \textit{k}. We choose an importance density $q(\bx_{n,k}^{j} \mid\bx_{n,k-1}^{}, \bY_{1:k},\psi_{k}^{j},\bx_{-n,k}^{j(l')})$ and then obtain the weight as
\begin{align}
 w_{n,k}^{ij} \propto  w_{n,k-1}^{i} \frac{p(\bY_{k} \mid \bx_{n,k}^{ij},\psi_{k}^{j},\bx_{-n,k}^{j(l')}) p(\bx_{n,k}^{ij} \mid \bY_{1:k-1},\psi_{k}^j,\bx_{-n,k}^{j(l')})}{q(\bx_{n,k}^{ij} \mid \bx_{n,k-1}^{i}, \bY_{1:k},\psi^j_{k}, \bx_{-n,k}^{j(l')})},
\end{align}
where $\bx_{n,k-1}^{i}$ denotes the $i$th particle sample after the selection criterion step at $k-1$. We choose the prior as the importance distribution, so that we have the following simplification, the standard bootstrap filter~\cite{Gordon1993}:
\begin{align}
 w_{n,k}^{ij} \propto w_{n,k-1}^{i} p(\bY_{k} \mid \bx_{n,k}^{ij},\psi_{k}^{j},\bx_{-n,k}^{j(l')}).
\end{align}
We normalise $w_{n,k}^{ij}$ to obtain $\widetilde{w}_{n,k}^{ij}$. We adopt a residual resampling step~\cite{Douc2005} to avoid the so-called degeneracy problem ~\cite{Doucet2000}. See Algorithm~\ref{alg:ipf} for the detailed description of this iPF.

\begin{algorithm}[!ht]
\small
\tcp{At time $k$ for the $n_k^{j}$th dipole of a total $N_k^{j}$ dipoles, at the $l'$th Gibbs sampling iteration with $I_k$ particles}
\For{$ i = 1,\dots,I_k$}{

    \tcp{Prediction}
    $\bullet$ Draw samples $\bx_{n,k}^{ij} \sim p(\bx_{n,k}^{j}|\bx_{n,k-1}^{i},\psi_{k}^{j},\bx_{-n,k}^{j(l')})$.\
    
    $\bullet$ Compute weights: $w_{n,k}^{ij} \propto w_{n,k-1}^{i} p(\bY_{k}|\bx_{n,k}^{ij},\psi_{k}^{j},\bx_{-n,k}^{j(l')})$.\
     }
    $\bullet$ Normalise weights $\tilde{w}_{n,k}^{ij} = \frac{w_{n,k}^{ij}}{\sum_i w_{n,k}^{ij}}$.
    
    \tcp{Resample}
    $\{\bx_{n,k}^{ij}, \tilde{w}_{n,k}^{ij}\}_{i =1}^{I_k}$ to $\{\bx_{n,k}^{(i^{'}j)}, \frac{1}{I_k}, i^{'}\}_{i^{'}=1}^{I_k}$.\
    
    \tcp{Gibbs iteration choice}
    $\bullet$ Update the $n_k^{j}$th state estimate $\bx_{n,k}^{j}$ by random selection from $\bx_{n,k}^{(i^{'}j)}$.  \\
    $\bullet$ Assign $\bx_{n,k}^{j(l')} = \bx_{n,k}^{j}$ in the Gibbs iteration.
\caption{Individual particle filter}
\label{alg:ipf}
\end{algorithm}

For the particle weight of the iPF in the birth move (a new target appears), the samples are drawn uniformly from the whole state space $\Omega$, and assigned equal weight $\frac{1}{I_k}$. For the death move (an existing target disappears), the corresponding particles of the selected target are deleted. 

As stated in Section \ref{Gibbs}, $\bx_{-n,k}^{j(l')}$ is then updated using the result from the current iPF run. Since it is impossible to obtain the ground-truth state $\bx_{n,k}^{j}$ for each $n_k^{j}$th dipole source directly, we randomly pick up the state estimate from the resampled state $\bx_{n,k}^{ij}$. When the particle weights at $k$ are not available, we randomly select a sample from $\bx_{n,k}^{i}$ according to the weight $\tilde{w}_{n,k-1}^{ij}$ from the previous time step. Once the current particle weights are obtained, we can obtain the estimate $\bx_{n,k}^{j}$ by randomly selecting a sample from the resampled particles $\{\bx_{n,k}^{(i^{'}j)} \}$ with an equal probability $p^{(i)} = \frac{1}{I_k}$, where $i^{'}$ represents the particle index after resampling. 

The main body of the proposed algorithm is described in Algorithm~\ref{alg:GMPF}. 

To sum up, we modify the original MPF algorithm in the following three ways: 

(1) We integrate an iterated Gibbs sampling procedure to generate the individual state estimate in each iPF. This enables us to obtain more reliable estimations in each $\bx_{-n,k}^{j}$ assisted iPF run. The number of iterations is controlled by the parameter $l$.

(2) Instead of dividing the state space into several subspaces, samples of each iPF are drawn from the same state space. In the dipole initialisation step, the samples are drawn from the $G$ vertices; they are propagated using the individual dipole dynamic model in $\Omega$. 

(3) Rather than using a weighted mean method, $\bx_{n,k}^{j}$ in the $n$th iPF is selected randomly from all the samples $\bx_{n,k}^{ij}$, where $i$ is the sample index. In practice, we randomly pick up one of the particle filter samples $\bx_{n,k}^{ij}$ with the equal probability $p^{(i)} = \frac{1}{I_k}$.

\subsection{Selection criterion scheme}
We then obtain three $N^{j}_k$ candidates and the corresponding states $\widetilde{\bX}_k^{j}$ from the GMPF. We apply the selection criterion scheme to find the optimal pair amongst all the available estimates. We can obtain the posterior probability as 
\begin{align}
p(N_k^j \mid \bY_{1:k}, \hat{N}_{1:k-1}) \propto p(\bY_k \mid \bY_{1:k-1}, \psi_k^j) p(N_k^j \mid \hat{N}_{k-1}).
\label{E:Bayes}
\end{align}
The estimate is:
\begin{align}
  \hat{N}_k
   = \arg \max p(\bY_k \mid \bY_{1:k-1}, \psi_k^j) p(N_k^j \mid \hat{N}_{k-1}),   
 \label{eq:HSC}
\end{align}
where $p(\bY_k \mid \bY_{1:k-1}, \psi_k^j) \approx \frac{1}{N_k^j} \sum_{i=1}^{I_k} \prod_{n=1}^{N_k^j} w_{n,k}^{ij}$. According to \cite{Doucet2000}, and assuming that the source posterior factorizes over $n$ (i.e., the sources are independent), we can then obtain $\hat{N}_k$ by selecting the $N^{j}_k$ with the highest probability. Then obtaining $\hat{\bX}_k$ from the corresponding posterior mean $\widetilde{\bX}_k^j$.

\begin{algorithm}[!htp]  
    Initialisation at $k=0$: compute $\overline{\bX}_0^{\bL}$, assign $N_0 = \overline{N}_0$. \\
    Randomly draw $\bX_0$ from $\Omega$ for each of the three cases.\\
    \For{$ k = 1,\dots,K$}{
    \tcp{Probabilistic ROI estimation}
        Compute $\overline{\bX}_k^{\bL}$ to obtain $\overline{N}_k$ and $\Psi_{k}$ (see Section \ref{ROISec}). \\
        \For{$j = (-), (0), (+)$ }{See Equation \eqref{dipolenumber}. \\
            Set $\bX_k^j = \hat{\bX}_{k-1}$.\\      
          \eIf{$j = (+)$}{\tcp{Birth move} 
                          Uniformly draw $I_k$ particles from $\Omega$ for the new dipole. \\
                          Calculate its state and append it to $\bX_{k}^{j}$.\\ 
                          }{\tcp{$j = (-)$ Death move}        
                          Randomly select a dipole estimate from $\bX_{k}^{j}$.
                          Delete the selected state and its corresponding $I_k$ particles.
                          }
          Initialise Gibbs step $\bX_k^{j(1)} = \bX_k^{j}$.  \\
          \For{$l' = 1, 2 \dots l$}{
          \tcp{Gibbs iteration}
          \For{$n = 1, \dots, N_k^{j}$}{
                 \tcp{iPF}
                  Follow Algorithm~\ref{alg:ipf} to obtain $\bx_{n,k}^{j(l')}$. \\
                  Update $\bX_k^{j(l')} = \{\bx_{1,k}^{j(l')}, \cdots, \bx_{n,k}^{j(l')}, \cdots, \bx_{N,k}^{j(l')} \}$. }
                  Set $\bX_k^{j(l'+1)} = \bX_{k}^{j(l')}$ \\
                  }
                  Assign $\bX_k^{j} = \bX_k^{j(l)}$ and obtain the pair $\{N_k^{j}, \widetilde{\bX}_k^{j}\}$.\\
        }
        \tcp{Selection criterion}
        Select $\{\hat{N}_k, \hat{\bX}_k\}$ from the three cases by Equation \eqref{eq:HSC}.\\
        Set the final estimates $\hat{N}_k$ and $\hat{\bX}_k$. \\
        \tcp{Adaptive filtering}
        Adjust $I_{k+1}$ w.r.t. evaluation result from Equation \eqref{eq:evaluate}. \\
    }
\caption{Bayesian multiple dipole localisation algorithm}
\label{alg:GMPF}
\end{algorithm} 

\subsection{Adaptive filtering} \label{adaptiveFilter}
As shown in Figure~\ref{Relationship}, the discrete amplitude matrix $\overline{\bX}^{\bL}_k$ from MNE is used in GMPF to assist the sampling procedure in every iPF run. The identified ROIs and their point sub-set $\Psi_{k}(n)$ are used to control the particle number and particle transition range. In practice, we calculate the localisation root mean squared error (RMSE) between the centre of each ROI and the dipole state estimation $\bX_{k}$ at each time step $k$. The localisation RMSE $e_k$ can be obtained from: 
\begin{align}
 e_k = D(\Psi_{k}(n),\bX_{k}),  \label{eq:evaluate}
\end{align}
where function $D(\cdot)$ computes the localisation RMSE between all elements in $\Psi_{k}(n)$ and $\bX_{k}$. The centre points of the ROIs $\bc^{n}_{k}$ are compared with the localisation estimation$\bX_{k}$. We compute the spatial distance $||\bX_{k} - \bc^{n}_{k}||$ between each of the the individual dipole state and the ROIs. We then obtain an $N_{k} \times \bg_k^{n}$ matrix $C_k$ that contains all pairs of the localisation RMSE. We execute the target association according to the RMSE matrix $C_k$. The targets in the state pairs with the smallest RMSE are then associated together. $e_k$ is then used as a reference criteria to adaptively adjust the particle number and the particle transition range in particle filtering at time $k+1$.

$e_k$ assists adaptive filtering mainly in two aspects: (1) the number of samples $I_{k+1}$ is modified with respect to the RMSE level $e_k$, a smaller $I_{k+1}$ is assigned when we obtain a lower $e_k$ and vice versa. (2) In the individual dipole propagation, the dipole dynamic range for the sample $\bx_k^{ij}$ depends on the value of $e_k$; a lower $e_k$ results in a smaller dipole dynamic range and vice versa. 

\section{Numerical results} \label{Results}

In this section, we present numerical results using the synthetic data. Since the ground-truth dipole location from real data remains unknown, the performance evaluation relies on the results from the synthetic data. 

\subsubsection{Simulation setting}
We adopted typical examples with both known and unknown number of dipoles. The number of dipoles varied between one and five. The ground-truth dipoles had unit amplitude in our simulation. The orientation of each dipole was set as normal to its corresponding sub-plane. Visualisations were carried out with tools further developed from those published in Helsinki BEM Library~\cite{Stenroos2007}. We generated the MEG data using a 204-magnetometer sensor setup. All magnetometers were distributed around the surface of the head. The state space $\Omega$ was strictly constrained within the pre-defined 1-layer real head cortex. The width of the brain was 136 mm in our simulation. According to empirical observations, a brain current source often appears and disappears in the same region, and the centre of the current source evolves within a small volume in the cortex. Therefore, it was reasonable for us to assume that all the dipoles were identical and independent of each other, and that each individual dipole might move within a pre-defined triangular sub-plane. We set the measurement SNR (signal to noise ratio) as 10. The measurement noise in the head model had a Gaussian distribution with zero mean and variance $\sigma_{\zeta}^2$, where $\sigma_{\zeta}$ was two times larger than that of the ground-truth noise. We tested each of the algorithms and the model with more than 30 repeated identical experiments. The number of dipoles for each iPF at $k=0$ was set as $I_0 = 10000$. 

In the simulation, we test both the GMPF algorithm with one Gibbs iteration (the original GMPF) and the GMPF algorithm with five Gibbs iterations. The GMPF algorithm with five Gibbs iterations outperforms the original GMPF algorithm in unknown dipole number examples, however it only provides slightly better results than that from the original GMPF algorithm in the known dipole number examples. So we only show the results from GMPF algorithm with five Gibbs iterations for the unknown dipole number case in Section \ref{unknown}. The term `GMPF' in the rest of this section refers to the original GMPF unless specified otherwise. 

In the remainder of this section, we compares the performance of different head models in examples with known and unknown numbers of dipoles. An example with five known numbers of dipoles is used to evaluate the localisation algorithms using the proposed continuous head model. For localisation with an unknown number of dipoles, we test and compare the performance of different particle filtering algorithms in an example with three and dynamic numbers of dipoles. We finally present an evaluation result particularly for the estimation of the dipole number.    

\subsection{localisation with known number of dipoles}
The performance of a known number of dipoles in terms of their localisation is easier and more accurate than the localisation with an unknown number of dipoles. In this section, we present two examples, one with three dipoles (two on the left and one on the right hemisphere) and other with five dipoles (two on the left and three on the right hemisphere). The three dipole example is used to compare three different head models.

\subsubsection{Head model comparison}
We compared the model performance between the spherical head model, discrete real head model, and the proposed continuous real head model. The spherical head model is a relatively old model that assumes the human head is a perfect spherical shape. This model was used in previous work ~\cite{Miao2013,Sorrentino2009}. The discrete head model is the 1-layer real head model generated using the BEM method, this model was used in ~\cite{Sorrentino2013,Chen2013A,Chen2013B}. In this paper we adopt the discrete model which contains $G = 8196$ discretised potential source points, all of which have fixed locations. A lead-field matrix with $G$ columns is then generated. We employ the NN interpolation method to convert the discrete model into the continuous model, which was presented in Section \ref{DataModel}. 

\begin{table}[!ht]
 \centering
 {\footnotesize
    \begin{tabular}{ | l | l | l | l | l | l | l | l | l | l |}
    \hline
    GTM / MM  & S/S   & S/D   & S/C   & D/S   & D/D   & D/C   & C/S   & C/D   & C/C\\ \hline
    SIR/2000  & 17.32 & 29.37 & 22.22 & 35.61 & 13.01 & 15.55 & 21.28 & 22.78 & 16.24\\ \hline
    MPF/2000  & 13.33 & 19.29 & 19.11 & 23.65 & 11.31 & 13.77 & 16.34 & 13.46  & \textbf{9.22}\\ \hline
    GMPF/2000 & 13.20 & 19.17 & 18.16 & 22.99 & 11.68 & 13.94 & 16.28 & 12.67  & \textbf{9.38}\\ \hline
    
    SIR/5000   & 15.28 & 28.22 & 21.35 & 35.06 & 11.37 & 13.10 & 19.49 & 17.44 & 15.36\\ \hline
    MPF/5000  & 11.57 & 16.32 & 18.06 & 21.39 & 10.73 & 12.89 & 16.45 & 12.08 & \textbf{7.10}\\ \hline
    GMPF/5000 & 11.77 & 16.28 & 18.24 & 20.12 & 10.25 & 12.45 & 16.99 & 11.54 & \textbf{7.53}\\ \hline
    
    SIR/10000  & 13.05 & 25.81 & 18.20 & 29.77 & 9.88  & 11.29 & 17.53 & 14.02 & 12.74\\ \hline
    MPF/10000 & 9.36  & 15.78 & 15.13 & 20.22 & 9.36  & 10.97 & 15.68 & 9.52  & \textbf{5.70}\\ \hline
    GMPF/10000 & 9.28  & 15.90 & 14.84 & 20.35 & 9.33  & 10.92 & 15.71 & 9.48  & \textbf{5.64}\\ \hline
    \end{tabular}}%
    \caption{RMSE comparison between three head models using three different particle filter algorithms. All units are in millimeters. GTM: the ground-truth model, MM : the measurement model, S: the spherical head model, D: the discrete head model, C: the continuous head model, SIR: the SIR particle filter, MPF: the multiple particle filter, GMPF: the the proposed Gibbs multiple particle filter, the number in the first column is the sample number employed by the particle filter.}
    \label{tabPD}
\end{table}

In Table~\ref{tabPD}, we present the numerical result with three dipoles with known locations. Three models are tested using a simple standard SIR particle filter, a standard MPF and the proposed algorithm. The entries with `S/S' (column entry) and `SIR' (row entry) corresponds to the method from ~\cite{Sorrentino2009}; the entries with `S/S' and `MPF' is similar to one of the methods adopted in ~\cite{Miao2013}; the entries with `D/D' and `MPF' was from our previous work ~\cite{Chen2013B}.

We vary the particle number from 2000 to 10000. The simulated data were generated and tested using all three models, for example, S/D in the table means that we generate the data using a spherical head model, and the measurement model we use in particle filtering is a discrete head model. 

\begin{figure}[!ht]
\centering
\includegraphics[width = 1\linewidth]{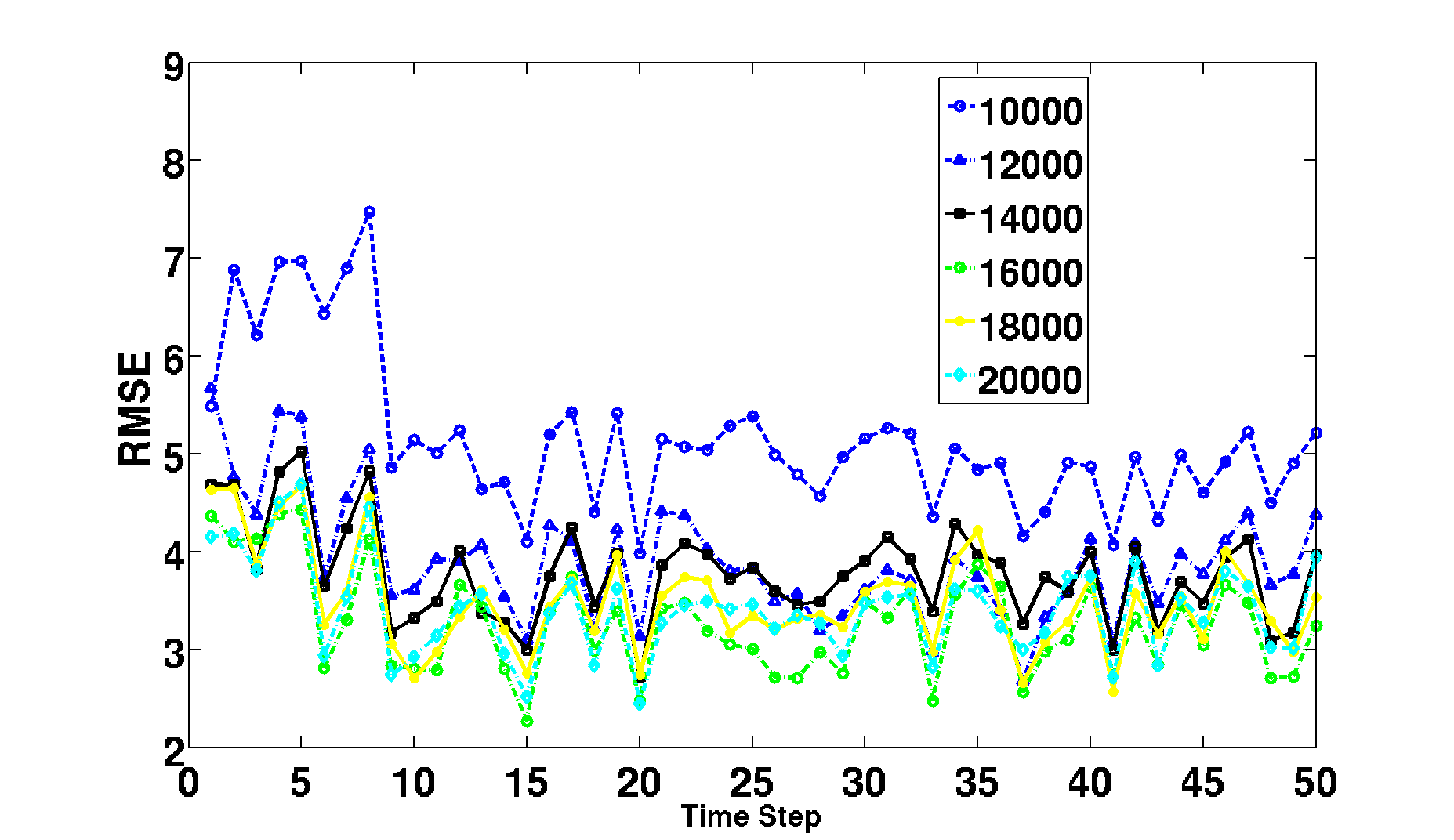}
\caption{RMSE performance using the C/C model pair and the GMPF algorithm with a varying particle number.}
\label{headmodel}
\end{figure}

Regarding the RMSE performance between different particle filtering algorithms, we can find that both MPF and GMPF perform better than the standard SIR. MPF and GMPF demonstrated similar performance. This is not surprising since the proposed GMPF algorithm is executed in a similar way to that of MPF in the fixed and known dipole number scenario. We also find that the RMSE performance improves with an increase in particle number. 

For those few entries with a similar performance between $I=2000$ and $I = 5000$ (e.g., column C/S of MPF and GMPF), a similar RMSE may occur due to model mismatching. In terms of head model comparison, it is as expected that S/S, D/D, and C/C achieve better performance. Amongst all model pairs for the same algorithm, the continuous head model performs better than the others, as shown in bold. We also find that the spherical model can only perform well when the ground-truth model is the same, while the discrete head model and the continuous head model are more robust to different data. 

In Figure~\ref{headmodel}, we show the detailed RMSE performance using the continuous head model and the proposed GMPF algorithm. For the initial particle numbers larger than 12000, the RMSE stays at the same level. This phenomenon is expected according to the individual dipole dynamic model, as a dipole only moves within the triangular sub-plane.  

\subsubsection{Known number of five dipoles}

\begin{figure}[!ht]
\centering
\subfigure[]{
\includegraphics[width = 0.48\linewidth]{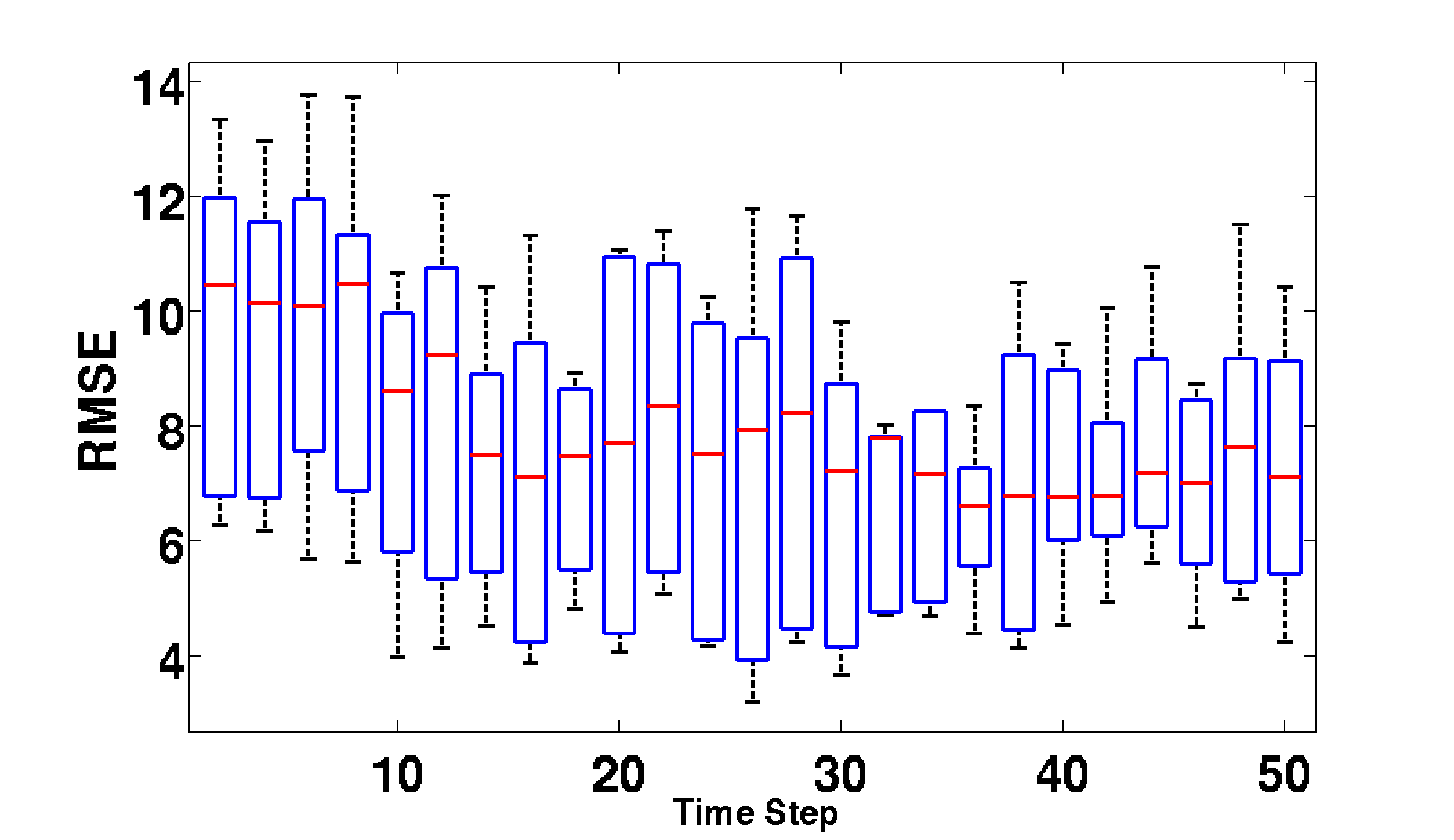}}
\subfigure[]{
\includegraphics[width = 0.48\linewidth]{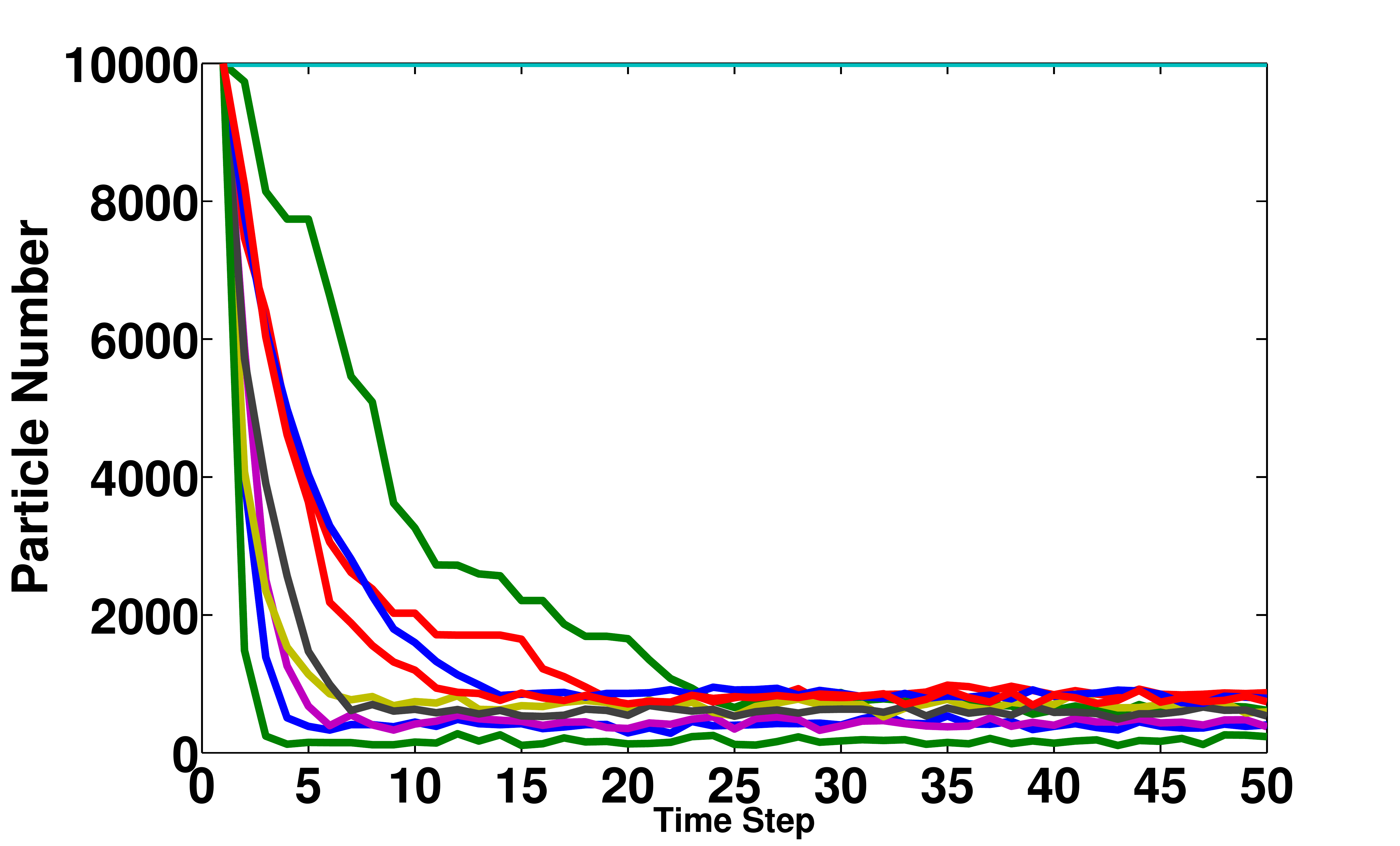}}
\subfigure[]{
\includegraphics[width = 0.48\linewidth]{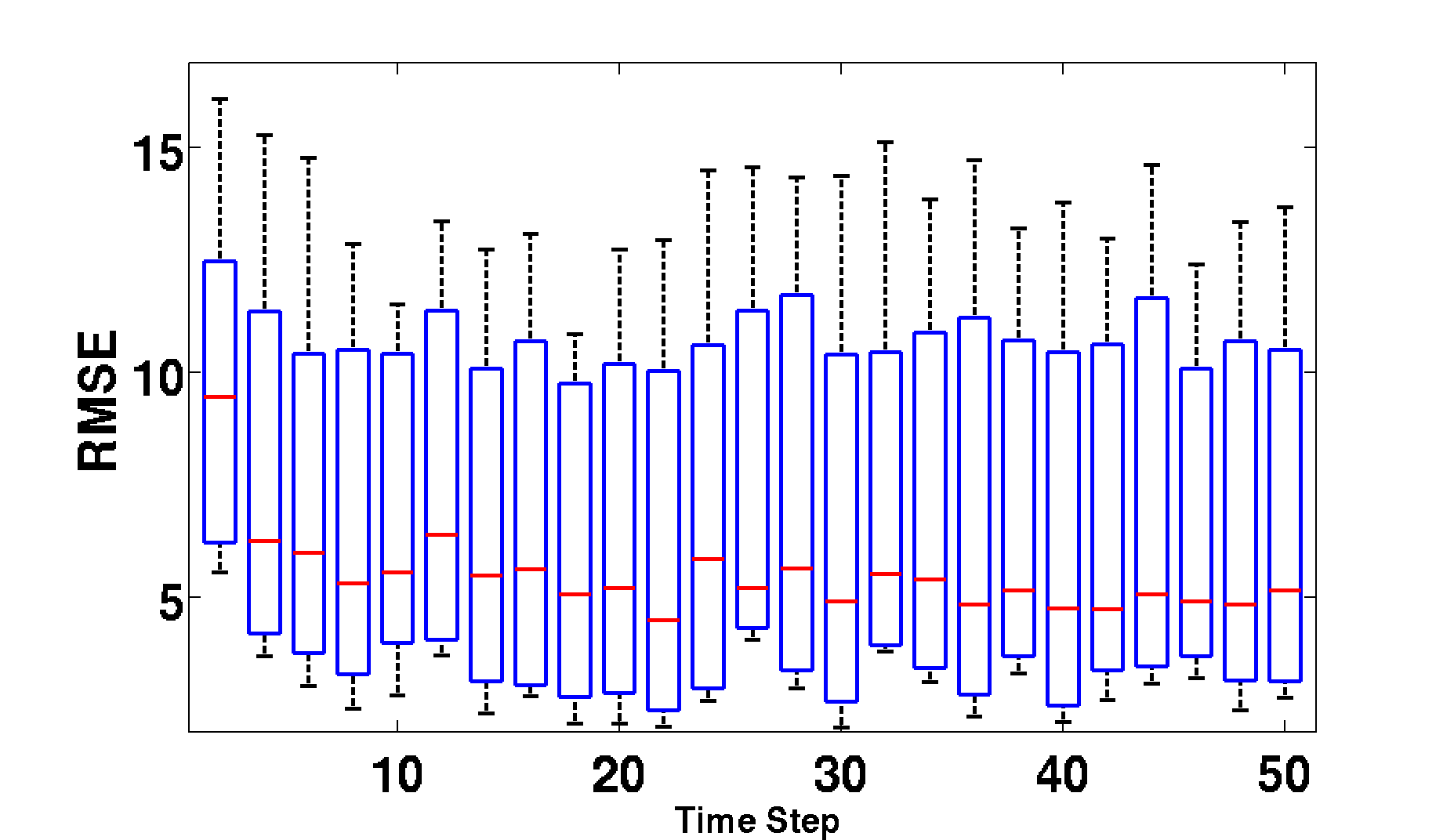}}
\subfigure[]{
\includegraphics[width = 0.48\linewidth]{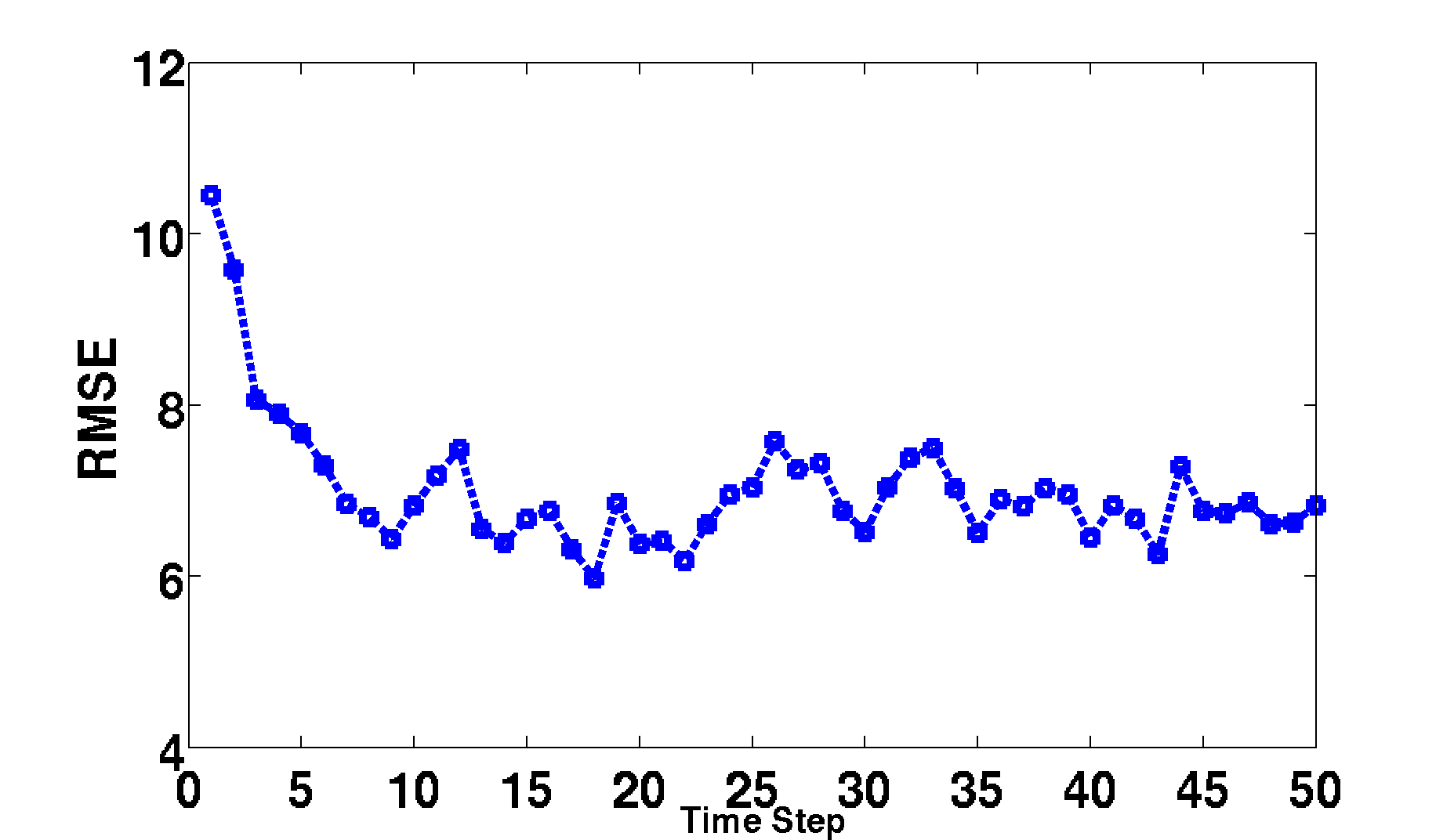}}
\caption{A five-dipole localisation example using the C/C model pair and the GMPF algorithm: (a) shows the average RMSE from the MPF algorithm. (b) shows a decreasing particle size in the algorithm run using the proposed adaptive filtering scheme. The different coloured lines represent the particle size of some identical trails. (c) shows the average RMSE from the GMPF algorithm. (d) shows the RMSE performance of an individual dipole in one of the trails.}
\label{K5}
\end{figure}

Figure~\ref{K5} shows the tracking results of five known dipoles using the continuous head model and the proposed algorithm. 

We can see from Figure~\ref{K5}(a) that the average RMSE over all five dipoles remains at the level of 8 mm. The box-plot ranges between 4 and 12 mm. In Figure~\ref{K5}(b) we show the changes of particle size for some of the identical trials. The initial particle size is $I_0 = 10000$. We can see that $I_k$ in most experiments  dramatically decreases to very low level from around $k = 5$. If we compare the two figures, one can observe that the proposed algorithm is able to achieve a good RMSE performance while adaptively eliminating the number of particles. This greatly saves the computational cost for multiple particle filter types of algorithms. The total particle number is $I_k$ multiplied by the number of dipoles. 

\begin{table}[!ht]
 \centering
 {\footnotesize
    \begin{tabular}{ | l | l | l | l | l | l | l | }
    \hline
    Alg / Dipole & 1 & 2 & 3 & 4 & 5 & Processing Time  \\ \hline
    MPF & 8.72 & 9.85 & 9.14 & 10.68 & 9.57 & 56 s \\ \hline
    GMPF & 7.02 & 9.33 & 7.58 & 10.24 & 9.17 & 28 s \\ \hline
    \end{tabular}}%
    \caption{A table comparison for the RMSE and the computer processing time, the particle size for both the MPF and the GMPF algorithms is 10000.}
    \label{tabK5}
\end{table}

From Figure~\ref{K5} (a) and (c), we can see that the average RMSE over five dipoles is almost at the same level for both MPF and GMPF. This can also be seen in Table~\ref{tabK5}. The middle five columns are the dipole index in the five dipole localisation example. This shows the average RMSE for each dipole over 30 identical experiments. We implemented the algorithms in Matlab using a computer with intel Core i7 CPU @ 3.7 GHz. The processing time for an iteration (with 50 time steps) of the MPF algorithm is 56 seconds and that of GMPF is 28 seconds, the total computational time for 30 iterations is around 1820 seconds for the MPF, and 1340 seconds for the proposed method. From the table we observe that the computational time for GMPF is less than that for MPF, this is due to the implementation of the adaptive filtering step in the proposed algorithm. Although we add an extra Gibbs updating step in GMPF, which increases the computational expense), the decreasing of particle number from adaptive filtering step still makes the GMPF faster than the MPF. 

\subsection{Localisation with unknown number of dipoles}
In real world applications, we are not able to obtain prior information of the dipole number, thus an example with an unknown dipole number needs to be tested to access the proposed model and algorithm. The algorithm is tested using two examples: an unknown three dipoles localisation problem and an unknown dynamic dipole numbers localisation. The dipole number in the latter example varies between three and five. The three unknown dipole number examples is used to evaluate the performance of the proposed algorithm. 

As we stated in Equation \eqref{dipolenumber}, the dipole number dynamics are modelled and estimated through $p(N_k \mid N_{k-1})$. In other words, the new estimated dipole number at time $k$ depends on the previous time step estimates $N_{k-1}$. For the initial dipole number at time $k = 0$ where there is no historical data available, we assign $N_k$ with the clustered MNE estimation result $\overline{N}_0$. 

\subsubsection{Evaluation of dipole number estimation}
The new dipole number estimates for $k > 0$ highly depend on the estimation at its previous time step. In our selection criterion step, there are three candidate pairs $\{N_k^{j}, \bX_k^{j}\}$ where $j = (-), (0) , (+)$ as stated previously. In this paper, we multiply the dipole number dynamic probability $p_k(\tau_{j})$ from Equation \eqref{dipolenumber} with the probability $p(N_k, \bX_k \mid \bY_{1:k})$ computed in the selection criterion step as the candidate posterior probability. We assign $p_k(\tau_{0}) = 0.5$, $p_k(\tau_{-}) = 0.25$ and $p_k(\tau_{+}) = 0.25$. The dipole number dynamic probability can be assigned using other designed schemes. Here we simply give the algorithm a prior knowledge (which comes from the observation using real examples) that the dipole number evolves slowly and it is more likely that the dipole number stays the same between two time steps. $N_k$ is assigned using the candidate pair with the highest normalised candidate posterior probability. 

\begin{figure}[!ht]
\centering
\subfigure[]{
\includegraphics[width = 0.48\linewidth]{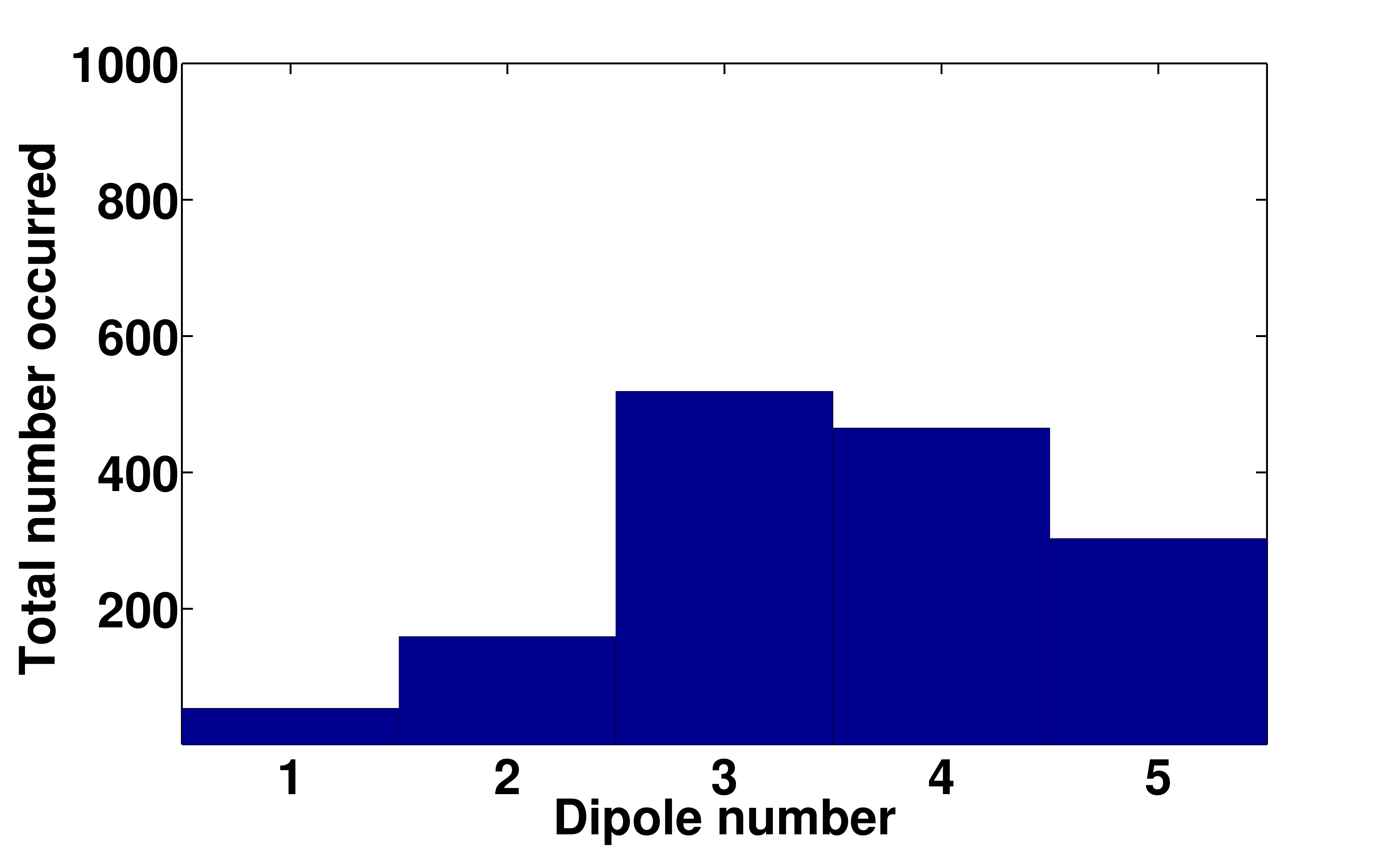}}
\subfigure[]{
\includegraphics[width = 0.48\linewidth]{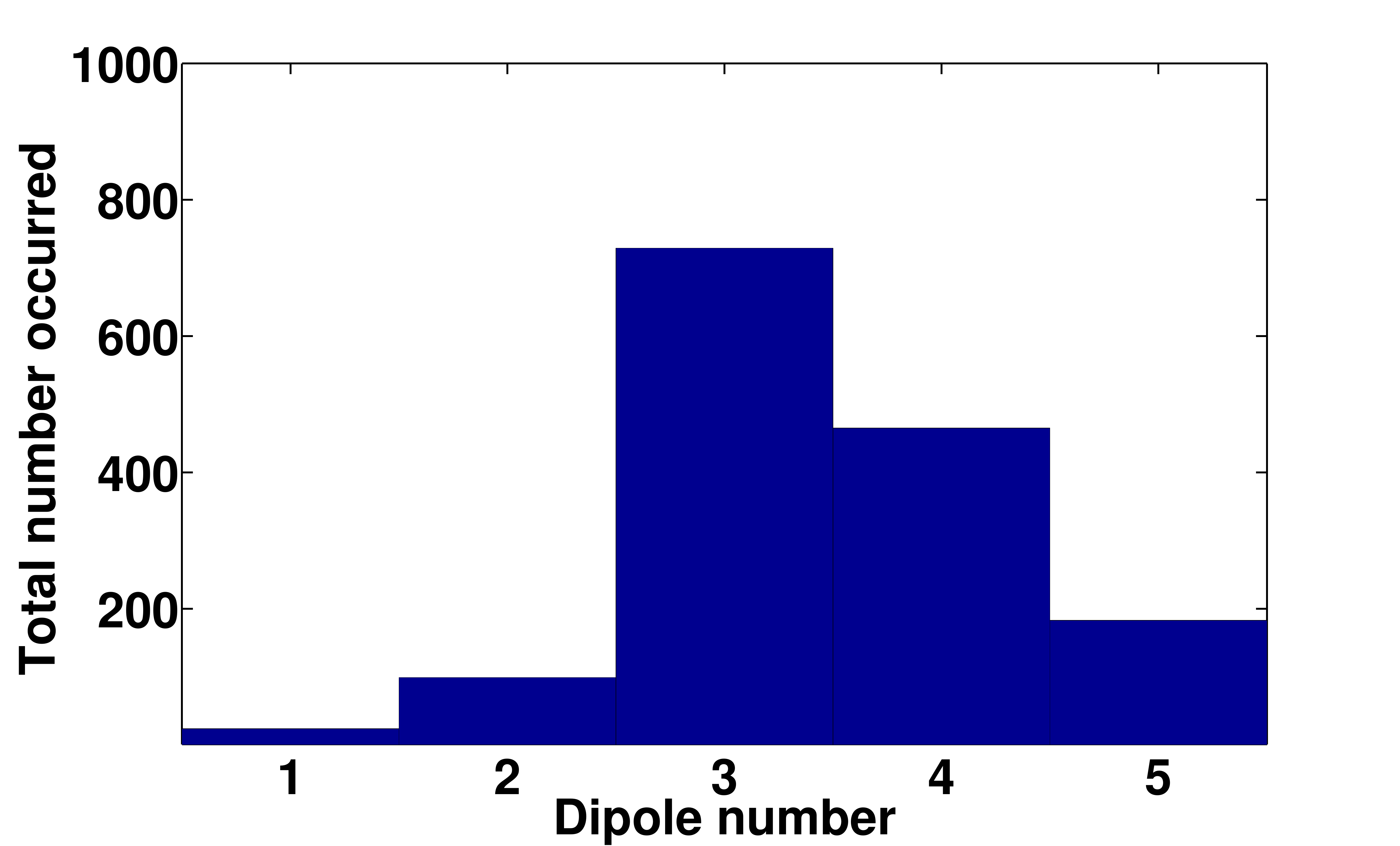}}
\subfigure[]{
\includegraphics[width = 0.48\linewidth]{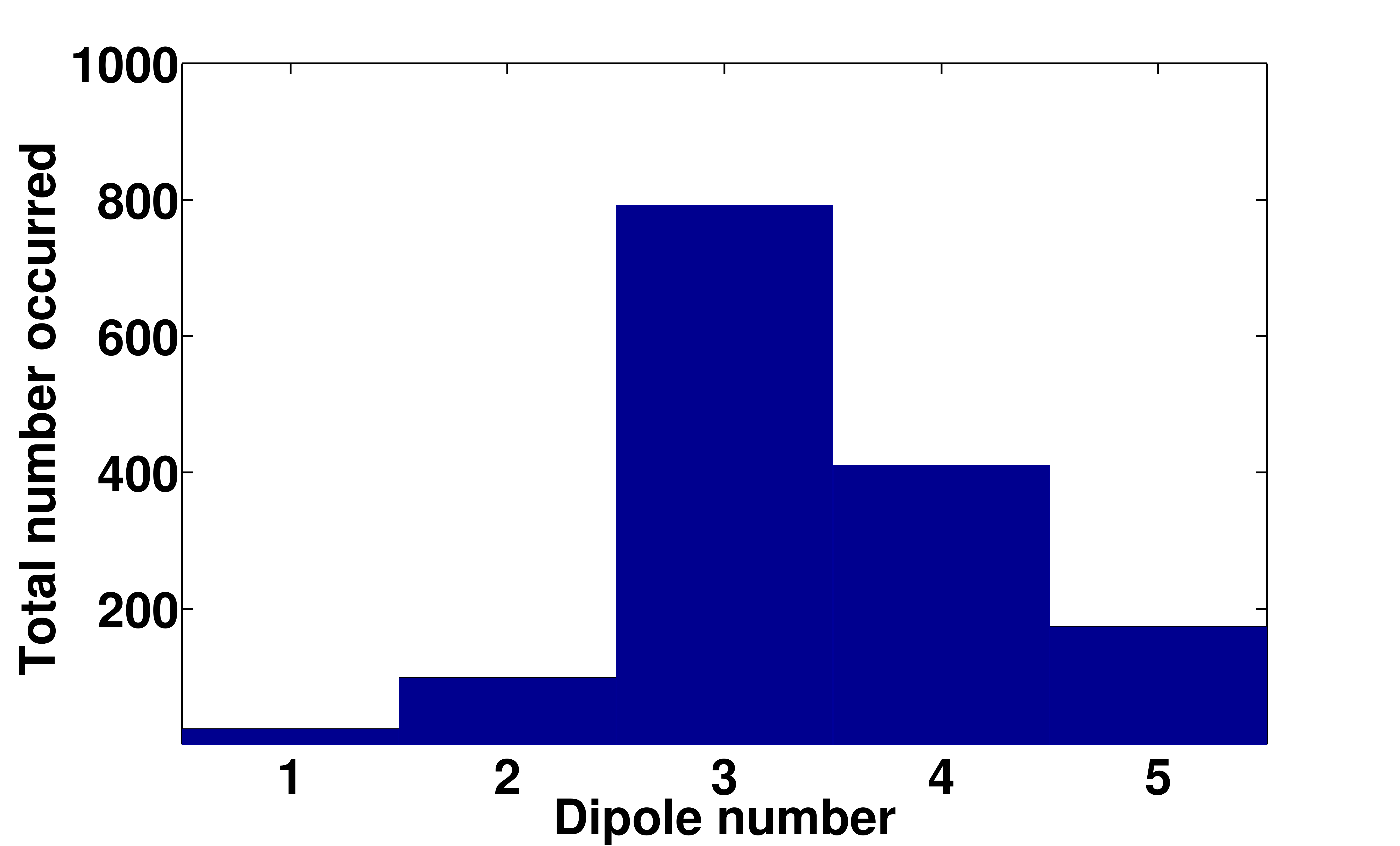}}
\subfigure[]{
\includegraphics[width = 0.48\linewidth]{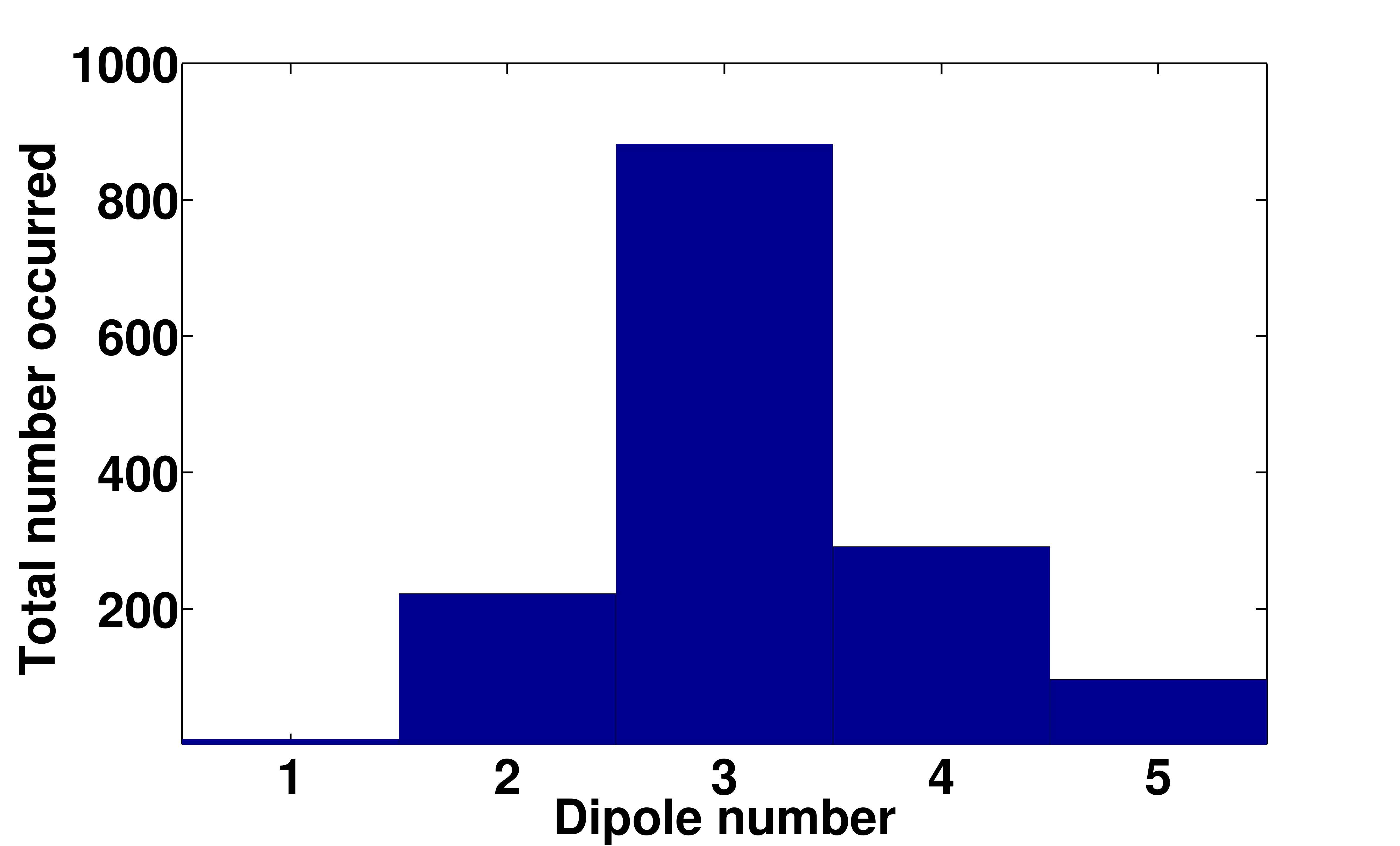}}
\subfigure[]{
\includegraphics[width = 0.48\linewidth]{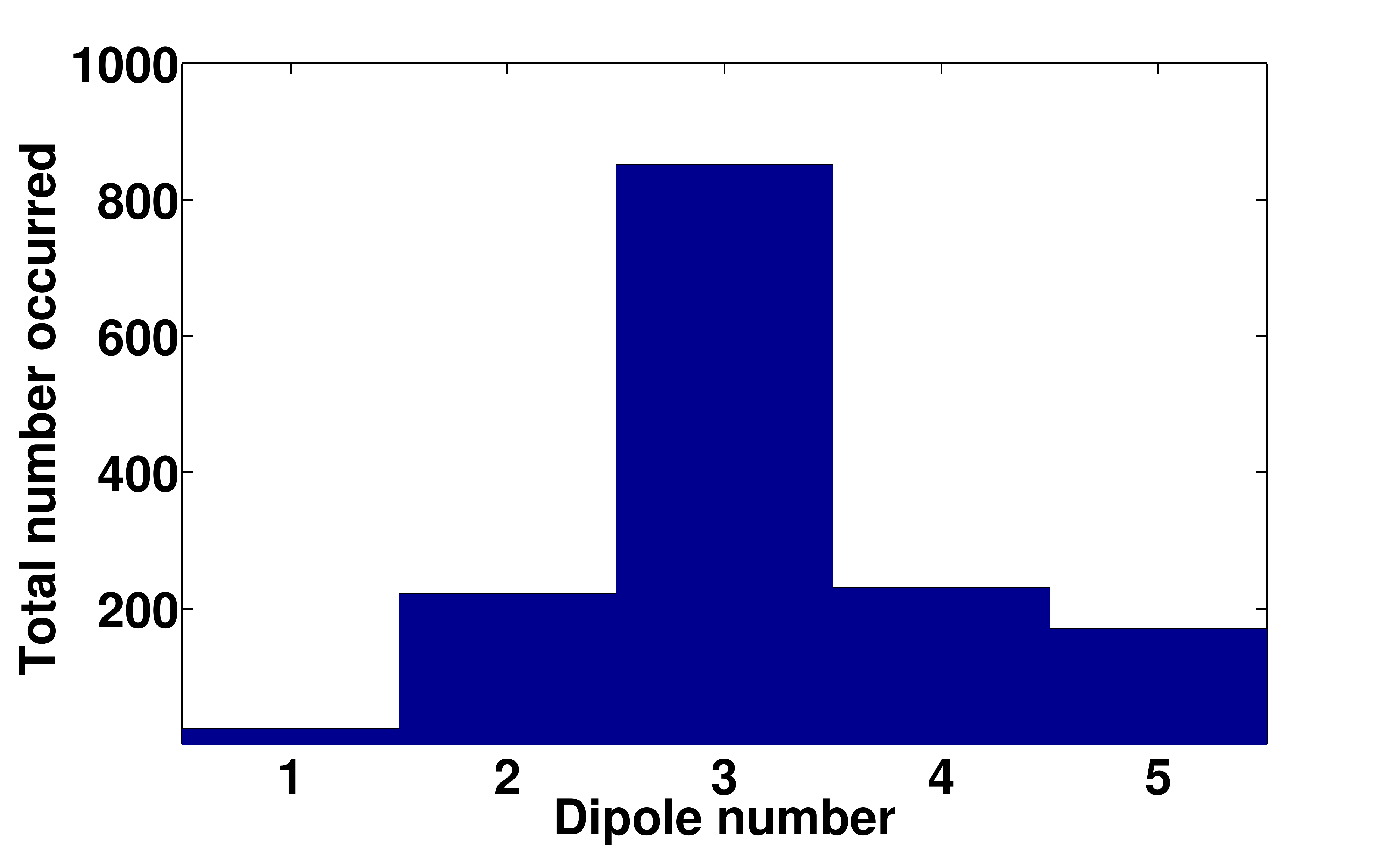}}
\subfigure[]{
\includegraphics[width = 0.48\linewidth]{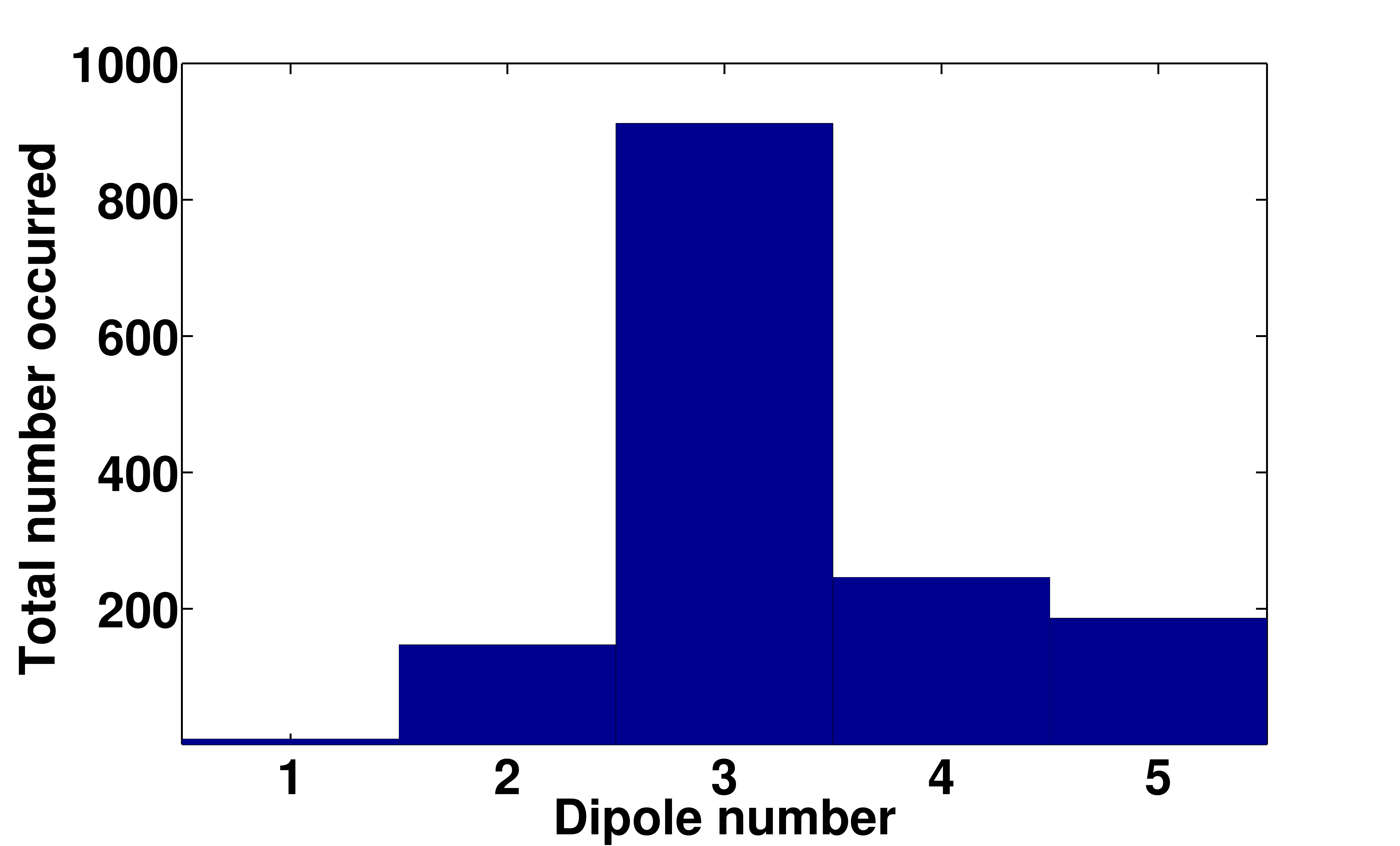}}
\caption{A histogram of the estimated dipole number over 30 identical trails with different particle size: (a) shows the performance using the MPF algorithm with 10000 particles, (b)-(f) shows the performance using the GMPF algorithm with 7000,10000,13000,16000 and 19000 particles, respectively.}
\label{U3}
\end{figure}

Here we use a three-dipole example to test the proposed approach. The example setting is the same as that in the known dipole example except that no prior dipole number knowledge is given to the algorithm. We apply both the MPF algorithm in paper \cite{Chen2013B} and the proposed GMPF algorithm to the example using a range of particle numbers. Figure \ref{U3} shows the histogram of the dipole number estimates. Each plot is the result from 30 identical experiments (with 50 time step length). Figure \ref{U3}(a) is the estimation result from the MPF algorithm in paper \cite{Chen2013B}, since its estimation does not depend on the state estimates from the particle filter, we only show one with 10000 particles. We apply our selection criterion assisted GMPF algorithm with a range of particle numbers. 

\begin{table}[!ht]
 \centering
 {\footnotesize
    \begin{tabular}{ | l | l | l | l | l | l | l | }
    \hline
    GMPF Particle Number & 7000 & 10000  & 13000 & 16000 & 19000 & MPF 10000  \\ \hline
    Avg. Dipole Number & 3.568 & 3.442 & 3.260 & 3.282 & 3.292 & 3.562 \\ \hline
    \end{tabular}}%
    \caption{Table shows the mean estimated dipole number using the GMPF algorithm with different particle size. The last column shows the performance from the MPF algorithm with 10000 particles.}
    \label{tabU3}
\end{table}

Table~\ref{tabU3} shows the average estimated dipole number over 30 iterations. Since the ground-truth dipole number is 3, we can observe that with an increase in the particle number from 7000 to 19000, the average estimated dipole number improves slightly. The numbers between GMPF 7000 and GMPF 10000 are similar; however if we check Figure \ref{U3}(a) and Figure \ref{U3}(b) for their corresponding histogram distribution, we can see there are more successful estimates (dipole estimate equal to 3) in GMPF 7000 than GMPF 10000. From the histogram we can also observe that, with an increase in the particle number using GMPF, the result does not improve much for particle sizes larger than 13000. 

\subsubsection{Dynamic unknown number of dipole example} \label{unknown}
We use a dynamic particle number setting in this example. The ground-truth dipole number varies between 3 and 5 during the 50 time steps. The number of dipoles is shown in the `True' row in Table~\ref{UD}. We use the MPF, the original GMPF and the GMPF with five Gibbs iterations to perform the localisation task; the particle numbers are all equals to 10000. 

In order to compare the algorithm performance in multiple target tracking scenario with the unknown number of targets, we adopt a multi-target performance evaluation method based on Optimal Subpattern Assignment Metric (OSPA) method~\cite{Ristic2011}. The method introduces a penalty term to punish the missed or false tracks when there is an unequal estimated dipole number to the ground-truth. We denote the ground-truth dipole number as $N_k^{\upsilon}$, and the ground-truth dipole state as $\bX_k^{\upsilon}$. For case $N_k \geq N_k^G$, the OSPA based distance $D_{p,c}(N_k^{\upsilon}, \bX_k^{\upsilon}, N_k, \bX_k)$ can be computed by:
\begin{align*}
 & D_{p,c}(N_k^{\upsilon}, \bX_k^{\upsilon}, N_k, \bX_k) \\
 & = [\frac{1}{N_k}(\min_{n_k} \sum_{n_k^{\upsilon} = 1}^{N_k^{\upsilon}}(\min(c,\|\bx_k - \bx_k^{\upsilon} \|_{p'}))^p + (N_k - N_k^{\upsilon})c^p)]^{\frac{1}{p}},
\end{align*}
where $p$ and $p'$ are the norm term, $c$ is the penalty term. In this paper, we adopt the value of p = p' = 1 and c = 20, to follow the instructions from ~\cite{Ristic2011}.   

\begin{figure}[!ht]
\centering
\subfigure[]{
\includegraphics[width = 0.48\linewidth]{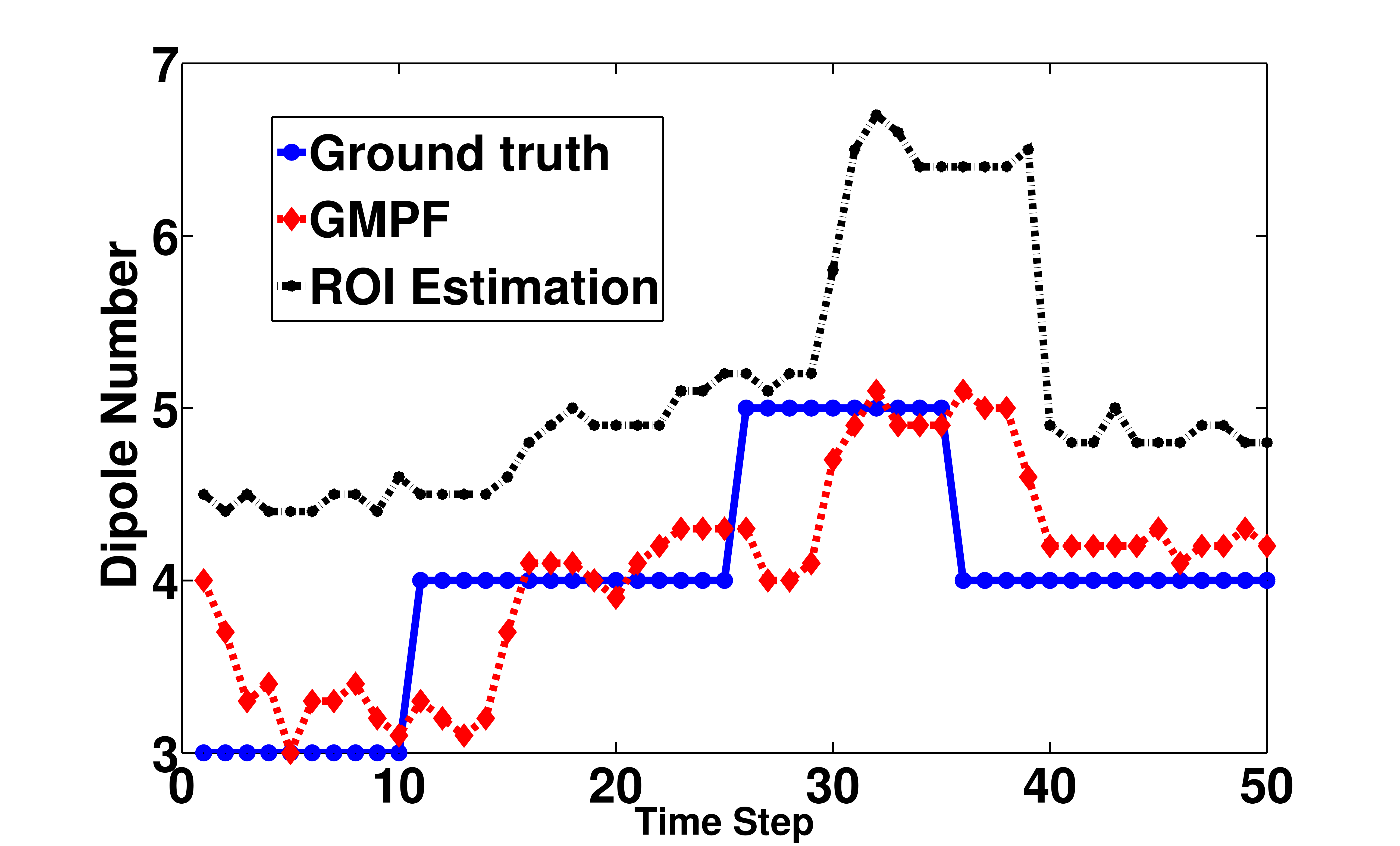}}
\subfigure[]{
\includegraphics[width = 0.48\linewidth]{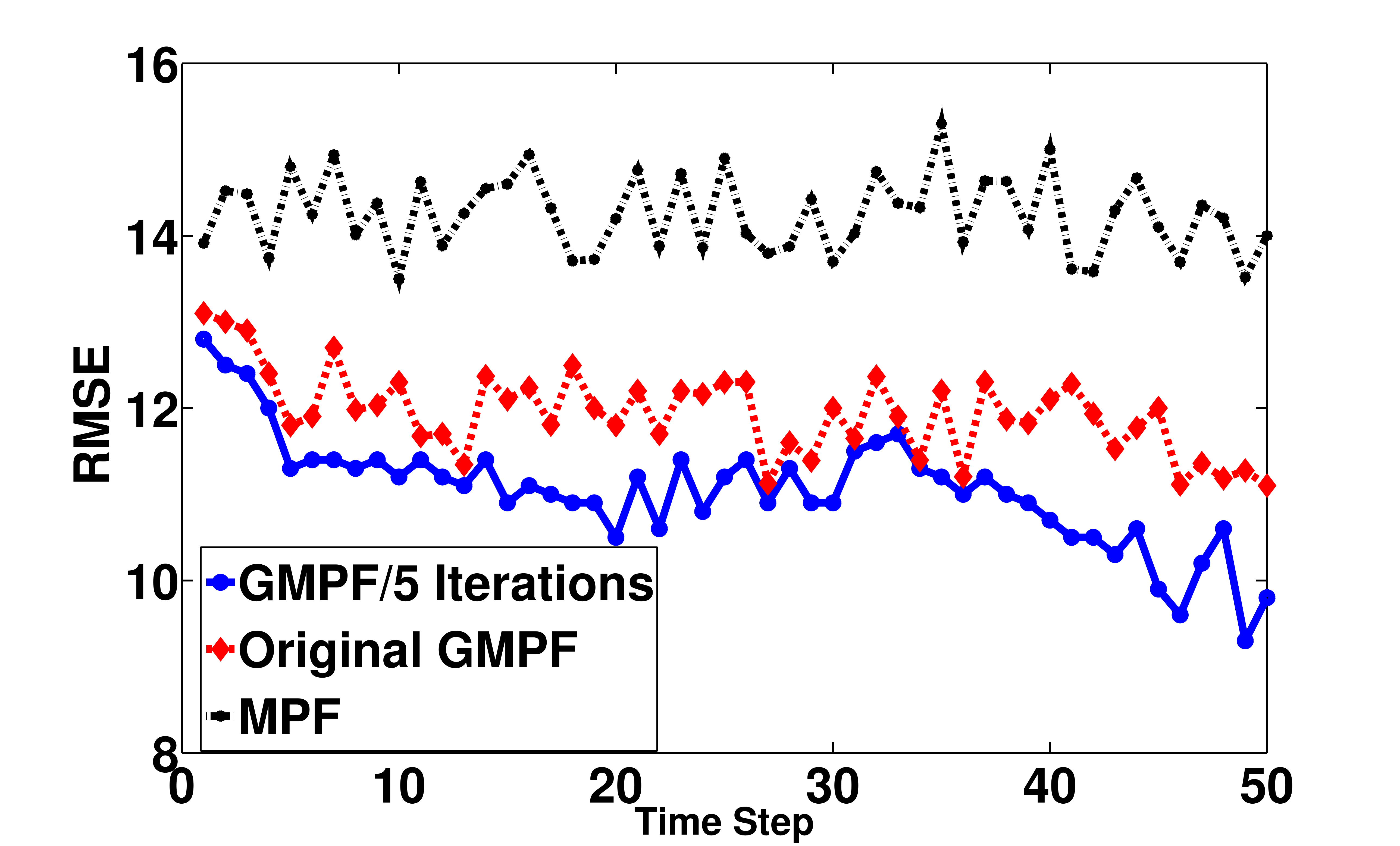}}
\caption{The estimated dipole number in the dynamic unknown number of dipole example. In (a), the red dashed line shows the mean estimated dipole number using the GMPF algorithm, the blue solid line is the ground-truth dipole number, the black dashed line is the number of the ROIs. (b) shows the localisation RMSE of the corresponding experiments, the blue solid line is the result using a GMPF algorithm with 5 times Gibbs iterations, the red dashed line is the one using a standard GMPF algorithm, and the black dashed line is the one using the MPF algorithm, all with the same particle size.}
\label{UDPLOT}
\end{figure}

In Figure \ref{UDPLOT}(a) we show the performance of the GMPF with five Gibbs iterations and the plot of the ground-truth dipole number (the blue solid line) versus the average estimated dipole number (the red dashed line) over 50 time steps. The black dashed line is the initial estimate from the probabilistic ROI estimation step. Figure~\ref{UDPLOT}(b) shows the corresponding average RMSE for localisation. 

We define the average RMSE larger than 30 mm as a lost track, thus the lost track percentages are $16.7\%$, $20\%$, $20\%$ for MPF, the original GMPF and the GMPF with five Gibbs iterations, respectively. The following RMSE calculations are based on the results with lost tracks removed. We find that the estimated dipole number follows that of the probabilistic ROI estimation at the first time step. 

\begin{table}[!ht]
 \centering
 {\footnotesize
    \begin{tabular}{ | l | l | l | l | l | l | l | l | l | l | l |}
    \hline
    Time $k$            & 5   & 10  & 15  & 20  & 25  & 30  & 35  & 40  & 45   & 50    \\ \hline
    True                & 3   & 3   & 4   & 4   & 4   & 5   & 5   & 4   & 4    & 4     \\ \hline
    ROIs                & 4.4 & 4.6 & 4.6 & 4.9 & 5.2 & 5.8 & 6.4 & 4.9 & 4.8  & 4.8   \\ \hline
    M/Avg $\hat{N}_k$   & 3.8 & 4.1 & 4.2 & 4.5 & 4.7 & 4.8 & 5.5 & 5.0 & 4.9  & 4.8   \\ \hline
    M/RMSE & 14.8 & 13.5 & 14.6 & 14.2 & 14.9 & 13.7 & 15.3 & 15.0 & 14.1 & 14.0 \\ \hline
    O/Avg $\hat{N}_k$   & 3.5 & 3.4 & 4.4 & 4.3 & 4.5 & 4.9 & 5.6 & 4.8 & 4.5  & 4.4   \\ \hline
    O/RMSE & 11.8 & 12.3 & 12.1 & 11.8 & 12.3 & 12.0 & 12.2 & 12.1 & 12.0 & 11.1 \\ \hline
    G/Avg $\hat{N}_k$    & 3 & 3.1 & 3.7 & 3.9 & 4.3 & 4.7 & 4.9 & 4.2 & 4.3  & 4.2   \\ \hline
    G/RMSE & 11.3 & 11.2 & 10.9 & 10.5 & 11.2 & 10.9 & 11.2 & 10.7 & 9.9 & 9.8 \\ \hline
    \end{tabular}}%
    \caption{Table for comparison between the ground-truth dipole number and the estimated dipole number (averaged over 30 iterations). ROIs represents the number of the ROIs estimated using the probabilistic ROI estimation scheme. The M/Avg $\hat{N}_k$ and the M/RMSE represent the mean estimated dipole number and the RMSE performance using the MPF algorithm, respectively. The O/Avg $\hat{N}_k$ and the O/RMSE represent the mean estimated dipole number and the RMSE performance using the original GMPF algorithm with no Gibbs iteration, respectively. The G/Avg $\hat{N}_k$ and the G/RMSE represent the mean estimated dipole number and the RMSE performance using the GMPF algorithm with 5 Gibbs iterations, respectively. All RMSE measures are in unit [mm].}
    \label{UD}
\end{table}

Table~\ref{UD} shows the numerical results of the same example for every 5 time steps. Both the estimation of the dipole number and the dipole localisation are acceptable given the standard deviation of the random walk equals to 4 mm. If we compare the results with that achieved by the known dipole number tracking example, the localisation error increases, and the average RMSE is approximately 2 times larger than that in the known dipole number tracking example. This is plausible since we have no prior knowledge of the dipole number. The Gibbs assisted GMPF algorithm has better dipole number estimation performance than the original GMPF with no Gibbs iteration. The RMSE performance for the GMPF with 5 Gibbs iterations is better than that of the original GMPF algorithm, as shown in the table. 

\section{Conclusion}
In this paper, we proposed a continuous real head model and a Bayesian particle filter algorithm to perform multiple dipolar sources localisation in MEG. The proposed algorithm integrated the probabilistic ROI estimation step and the selection criterion step to assist the estimation of the dipole number and the adjustment of the particle size. 

The dipole number estimation in the multiple dipole tracking problem remains a difficult issue. Dimensionality is another issue that needs to be considered. We adopted the GMPF scheme to avoid the dimensionality problem by assigning each target an iPF. We also incorporated the probabilistic ROI estimation step to provide the tracking step with prior information.

However, this method highly depends on the deterministic estimates at the initial step from the MNE. It does not perform well when the dipole number dramatically changes in a short time period. In order to address this problem, we modified the deterministic probabilistic ROI estimation step to a partly probabilistic estimation: we modified the dipole number dynamic model and allowed three potential dipole number guesses. The three candidate pairs are compared and we employed selection criterion scheme to select the pair with the highest probability. We also added Gibbs sampling step to assist tracking in the multiple particle filter. As we have shown in the numerical results, this approach achieves better localisation and dipole number estimation performance than the previously proposed algorithms.

The computational cost of an MPF algorithm approximately equals to the product of a single particle filter multiplied by the number of identified dipoles. However, it is not necessary to keep a large volume of particles for every time step when the tracking estimate is close enough to the ground-truth. We proposed the adaptive filtering scheme to reduce the computational cost by diminishing particle number when the localisation error is negligible. 

We are currently working on a fully probabilistic dipole number estimation method and we will incorporate it into our current localisation algorithm. As the brain current source is a continuous state space in reality, we will explore a better continuous modelling scheme to interpret the brain's current sources. Further research will also focus on decreasing the computational time. This may be addressed by using the parallel computing technique and by implementing a Rao--Blackwelisation method in the linear part of the model. 

\section*{Acknowledgment}
The authors would like to thank Professor Rik Henson and Dr. Olaf Hauk from the MRC Cognition and Brain Sciences Unit in Cambridge, UK, and Dr. Matti Stenroos from Aalto University in Finland, for fruitful discussions and for providing data for the experiments.

%\vfill\pagebreak

\section{Reference}
\bibliographystyle{IEEEbib}
\bibliography{references}

\end{document}